\documentclass[twocolumn,showpacs,preprintnumbers,amsmath,amssymb]{revtex4}
\usepackage[latin1]{inputenc}
\usepackage{amsmath}
\usepackage{amssymb,mathrsfs}
\usepackage{graphics}
\usepackage{epsfig}
\newtheorem{definition}{Definition}
\newtheorem{proposition}{Proposition}
\newtheorem{theorem}{Theorem}

\newtheorem{corollary}{Corollary}
\newcounter{figurecounter}
\newcounter{stepcounter}
\newcounter{asscounter}

\newenvironment{assumption}{\refstepcounter{asscounter}\bf Assumption \theasscounter:\it }{\medskip}

\newcommand{\ignore}[1]{}

\newcommand{\e}{{\rm e}}
\newcommand{\I}{{\rm i}}

\newcommand{\veps}{\varepsilon}

\newcommand{\NNN}{\mathbb{N}}
\newcommand{\RRR}{\mathbb{R}}

\newcommand{\CCC}{\mathbb{C}}

\newcommand{\A}{{\cal A}}
\newcommand{\B}{{\cal B}}
\newcommand{\C}{{\cal C}}
\newcommand{\D}{{\cal D}}

\renewcommand{\H}{{\cal H}}

\renewcommand{\L}{{\cal L}}
\newcommand{\M}{{\cal M}}

\renewcommand{\O}{{\cal O}}
\renewcommand{\P}{{\cal P}}
\newcommand{\R}{{\cal R}}

\newcommand{\W}{{\cal W}}

\begin{document}

\title{\vspace{-2cm}Constrained Quantum Systems as an Adiabatic Problem} 
\thanks{Supported by the DFG within the SFB/Transregio 71.  }

\author{Jakob Wachsmuth}
\email{jakob@ipke.de} 

\author{Stefan Teufel}
\email{stefan.teufel@uni-tuebingen.de. }
\affiliation{University of T\"ubingen, Institute of Mathematics, Auf der Morgenstelle 10, 72076 T\"ubingen, Germany.}



\begin{abstract}
We derive the effective Hamiltonian for a quantum system constrained to a submanifold (the constraint manifold) of configuration space (the ambient space) in the asymptotic limit where the restoring forces tend to infinity. In contrast to   earlier works   we consider at the same time the effects of variations in the constraining potential and the effects of interior and exterior geometry which appear at different energy scales and thus provide, for the first time, a complete picture ranging over all interesting energy scales. We show that the leading order contribution to the effective Hamiltonian is the adiabatic potential given by an eigenvalue of the confining potential well-known in the context of adiabatic quantum wave guides. At next to leading order we see effects from the variation of the normal eigenfunctions in form of a Berry connection. We apply our results to quantum wave guides and provide an example for the occurrence of a topological phase due to the geometry of a quantum wave circuit, i.e.\ a closed quantum wave guide.
\end{abstract} 
 
\maketitle

 
\section{Introduction}

The derivation of effective Hamiltonians for constrained quantum systems has been considered many times in the literature with different motivations and  applications in mind. Roughly speaking, the available results split into two   different categories which are related to two different energy scales. In the context of adiabatic quantum wave guides one considers the situation where the strong forces restricting the particle to the wave guide change their form along the direction of propagation. The eigenvalues of the transverse Hamiltonian thus also vary along this direction  and produce an effective adiabatic potential for the tangential dynamics, i.e.\ for the propagation. In this case the tangential kinetic energy   is of the same order of magnitude as the energy in the transversal modes.
The geometry of the wave guide plays no role at this level.  On the other hand, in the literature concerned primarily with the effects of the geometry of constraint manifolds \cite{FrH,Ma1,Mit} on the effective Hamiltonian, it is  assumed  that the constraining forces are ``constant'' along the constraint manifold. This is because the geometric effects are much smaller and would be dominated by the adiabatic potential otherwise. It is thus assumed that the tangential kinetic energy    is of the same small magnitude as the geometric effects  and thus much smaller than the transversal energies.  

In this paper we show how these two regimes are related and derive an effective Hamiltonian valid on all interesting energy scales. It contains contributions from the adiabatic potential, from a generalized Berry connection and from the intrinsic and extrinsic geometry of the constraint manifold. The derivation is based on super-adiabatic perturbation theory and a mathematically rigorous treatment of the problem is given in \cite{WT}. 
We present our results first on a general and abstract level. However, there are several concrete applications which have motivated us and the many predecessor works, most notably molecular dynamics and adiabatic quantum wave guides. In Section~\ref{waveguides}  we apply our results to adiabatic quantum wave guides and, in particular, obtain new results about global geometric effects in quantum wave circuits, i.e.\ closed wave guides.

\subsection{Qualitative discussion of the   results}

Although the mathematical structure of the linear Schr\"odinger equation
\begin{equation}\label{SE1}
\I \partial_t \Psi = - \Delta  \Psi + V\Psi =: H\psi \,,\quad \Psi|_{t=0}\in L^2(\mathcal{A}, d\tau)\,,
\end{equation}
is quite simple, in many cases the high dimension of the underlying configuration space   $\mathcal{A}$
makes even a numerical solution  impossible. Therefore it is important to identify situations where the dimension can be reduced by approximating the solutions of the original equation~(\ref{SE1}) on the high dimensional configuration space  $\mathcal{A}$ by solutions of an \textit{effective equation} 
\begin{equation}\label{SE2}
\I \partial_t \psi = H_{\rm eff}\psi \,,\qquad \psi|_{t=0}\in L^2(\mathcal{C}, d\mu)\otimes\CCC^M\,,
\end{equation}
on a lower dimensional configuration space $\mathcal{C}$. The factor $\CCC^M$ allows for the possibility of additional internal degrees of freedom in the effective description.

\smallskip

A  famous example for such a reduction is the time-dependent Born-Oppenheimer approximation: Due to the small ratio $\frac{m_{\rm el}}{m_{\rm nu}}$ of the mass  $m_{\rm el}$ of an electron and the mass $m_{\rm nu}$ of a typical nucleus, the molecular Schr\"odinger equation 
\begin{eqnarray}\label{BO}
\I \partial_t \psi &=& -\tfrac{1}{m_{\rm nu}}\Delta_{x} \Psi -\tfrac{1}{m_{\rm el}}\Delta_{y}  \Psi +V \Psi,\\
&&\qquad\qquad\Psi|_{t=0}  \in  L^2(\RRR^{3(n+m)}, dx\,dy),\nonumber
\end{eqnarray}
on the full configurations space $\RRR^{3(n+m)}\hat = \,\mathcal{A}$ of electrons and nuclei, may be approximated by an equation  
\begin{equation*}
\I \partial_t \psi \,=\, - \,\tfrac{1}{m_{\rm nu}}\Delta_{x}  \psi + E_{\rm el}\psi\,,\quad \psi|_{t=0}\in L^2(\RRR^{3n}, dx)\,,
\end{equation*}
on the lower dimensional configuration space $\RRR^{3n} \hat = \,\mathcal{C}$ of the nuclei only. In this case
 the interaction $V (x,y)$ of all particles is replaced by an electronic energy surface $E_{\rm el}(x)$, which serves as an effective potential for the dynamics of the nuclei. 
 The assumption here is   that the electrons   remain in an eigenstate of the electronic Hamiltonian $H_{\rm e}(x) = -\tfrac{1}{m_{\rm el}}\Delta_{y}  +V (x,y)$ corresponding to the eigenvalue $E_{\rm el}(x)$. This assumption is typically satisfied, since the light electrons move fast compared to the heavy nuclei and thus the electronic state adjusts adiabatically to the slow motion of the nuclei. This is an example  of \textit{adiabatic decoupling} where the reduction in the size of the effective configuration space stems from different masses in the system.

\smallskip

A physically different but mathematically similar  situation where such a dimensional reduction is possible are constrained mechanical systems. In these systems   strong forces   effectively constrain the system to remain in the vicinity of a submanifold $\mathcal{C}$ of the configuration space $\mathcal{A}$. 

For classical Hamiltonian systems on a Riemannian manifold $(\A,G)$ there is a straight forward mathematical reduction procedure. 
One just restricts the Hamilton function  to $\C$'s cotangent bundle $T^*\C$ by embedding $T^*\C$ into $T^*\A$ via the metric~$G$ and then studies the induced dynamics on $T^*\C$. 
For  quantum systems Dirac \cite{D} proposed to quantize the restricted classical Hamiltonian system on the submanifold following an 'intrinsic' quantization procedure. However, for curved submani\-folds $\mathcal{C}$ there is no unique quantization procedure. One natural guess would be
 an effective Hamiltonian   $H_{\rm eff}$ in (\ref{SE2}) of the form
\begin{equation}\label{Heff0}
H_{\rm eff} = -\Delta_{\mathcal{C}} + V|_\mathcal{C}\,,
\end{equation}
where $\Delta_\mathcal{C}$ is the Laplace-Beltrami operator on $\mathcal{C}$ with respect to the induced metric and $V|_\mathcal{C}$ is the restriction of the potential $V:\mathcal{A}\to \RRR$ to $\mathcal{C}$. 

To justify or invalidate the above procedures from first principles, one needs to model  the constraining forces within the   dynamics (\ref{SE1}) on the full space $\mathcal{A}$. This is done by adding a  localizing part to the potential $V$.  Then one analyzes the behavior of solutions of (\ref{SE1}) in the asymptotic limit where the constraining forces become very strong and tries to extract a limiting equation on~$\mathcal{C}$. This limit of strong confining forces has been studied in  classical mechanics and in quantum mechanics many times in the literature. 

The classical case was first investigated by Rubin and Ungar \cite{RU}, who found that the effective Hamiltonian for the motion on the constrained manifold contains an extra potential   that accounts for the energy contained in the normal oscillations. The quantum mechanical analouge of this extra potential is the adiabatic potential. The intrinsic geometry of the submanifold only appears in the definition of the kinetic energy $\frac{1}{2}g(p,p)$, its embedding into the ambient space $\mathcal{A}$ plays no role.

On the other hand, for the quantum mechanical case Marcus \cite{Mar} and later on Jensen and Koppe~\cite{JK} and Da Costa \cite{DC} pointed out that  the limiting quantum Hamiltonian contains a potential term, the geometric potential,  that  depends on the embedding of the submanifold~$\mathcal{C}$ into the ambient space~$\mathcal{A}$. 
But these statements (like the more refined results by Froese-Herbst \cite{FrH}, Maraner \cite{Ma1} and Mitchell \cite{Mit}) require  that the constraining potential is the same at each point on the constraint manifold. The reason behind this assumption is that  in the limit of strong confinement the adiabatic potential is much larger (by two orders in the adiabatic parameter) than the geometric potential. For the geometric potential to be of leading order one must thus assume that the tangential kinetic energy is of the same small order. Then one ends up in the situation where the energy in the transversal modes is much larger than the typical tangential energies and where, by assumption, any transfer of energy between transversal and tangential modes is suppressed.
In conclusion, the effective Hamiltonian obtained in this way   describes the constrained system only for very small energies and under very restrictive assumptions on the confining potential. Note that in many important applications the assumption of a constant confining potential is violated. For example for the reaction paths of molecular reactions, the valleys vary in shape depending on the configuration of the nuclei.  

In this work we present   a  general result concerning the precise form of the limiting dynamics~(\ref{SE2}) on an arbitrary constraint manifold $\mathcal{C}$ starting from (\ref{SE1}) on the ambient space $\mathcal{A}$ with a strongly confining potential $V$. 
The  most important new aspect of our result is that we allow for confining potentials that vary in shape and for solutions with normal and tangential energies of the same order and, at the same time, capture the effects of   geometry. As a consequence, our effective Hamiltonian on the constraint manifold has a richer structure than earlier results and resembles, at leading order, the results from classical mechanics. However, similar to the hierarchic structure of the spectrum of molecules, with electronic, vibrational and rotational levels, now different geometric effects appear in the higher order corrections to the effective Hamiltonian. 
We note that in the limit of small tangential energies and under the same restrictive assumptions on the confining potential we recover the limiting dynamics by Mitchell~\cite{Mit}.

The key observation for our analysis is that the problem is an adiabatic limit and has, at least locally, a structure similar to the Born-Oppenheimer approximation in molecular dynamics. In particular, we transfer ideas from adiabatic perturbation theory, which were developed on a rigorous level by Nenciu-Martinez-Sordoni \cite{MS,NS,So} and Panati-Spohn-Teufel in \cite{PST1,PST2,Te} and independently on a theoretical physics level by Belov-Dobrokhotov-Tudorovskiy in \cite{BDT}, to a non-flat geometry. We note that the adiabatic nature of the problem  was observed many times before in the physics literature, e.g.\ in the context of adiabatic quantum wave guides and thin films~\cite{BDT2,BDNT}. But the only work considering  constraint manifolds with general geometries in quantum mechanics from this point of view so far is \cite{DT}, where only the leading order dynamics of localized semiclassical wave packets is analyzed and effects of geometry or geometric phases play no role.  We thus believe that our effective equations have not been derived  before, neither on a mathematical nor on a theoretical physics level.

\subsection{The scaling explained in a simple example}\label{SubSecScaling}

\begin{figure}[b] 
\includegraphics[width=8.5 cm]{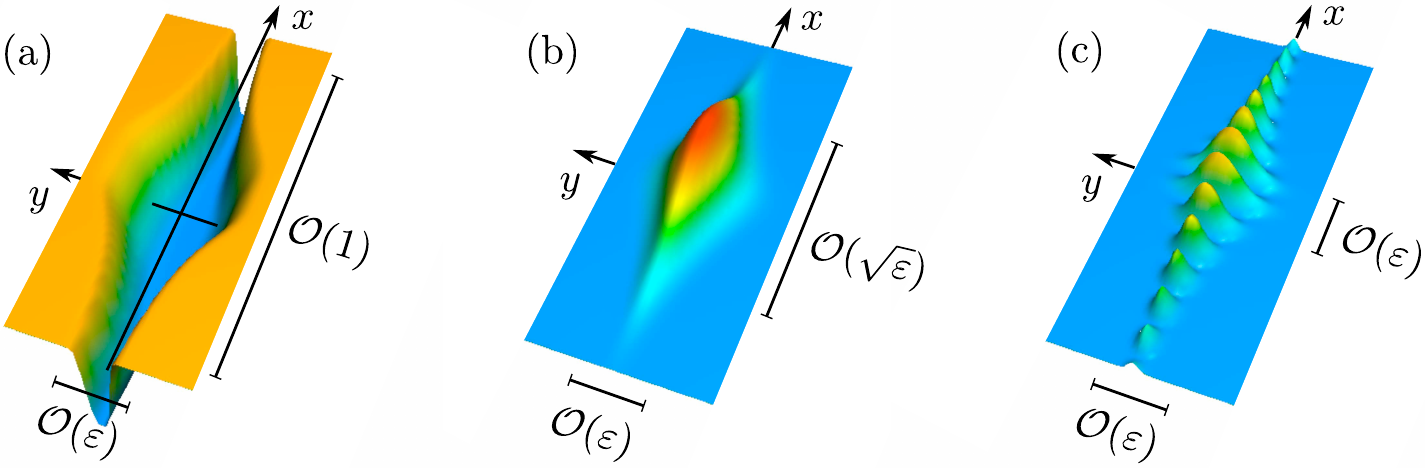}
 \caption{\label{fig1}In (a) we plotted a potential for a waveguide which widens near $x=0$. The widening lowers the energy of normal modes and thus produces an attractive effective potential for the motion in $x$-direction. In (b) the modulus of the ground state wave function is sketched. Its variation in $x$-direction is slower than in $y$-direction, but its tangential derivatives already grow in $\veps$. In (c) the modulus of an excited state with energy of order one above the ground state is sketched. Its variation in $x$-direction is on the same scale as the confinement, i.e.\ it oscillates on a scale of order $\veps$. Thus any analysis assuming bounded tangential derivatives of the solutions will be restricted to confining potentials with constant profile.}
\end{figure}

\begin{figure}[t] 
\includegraphics[width=7.5 cm]{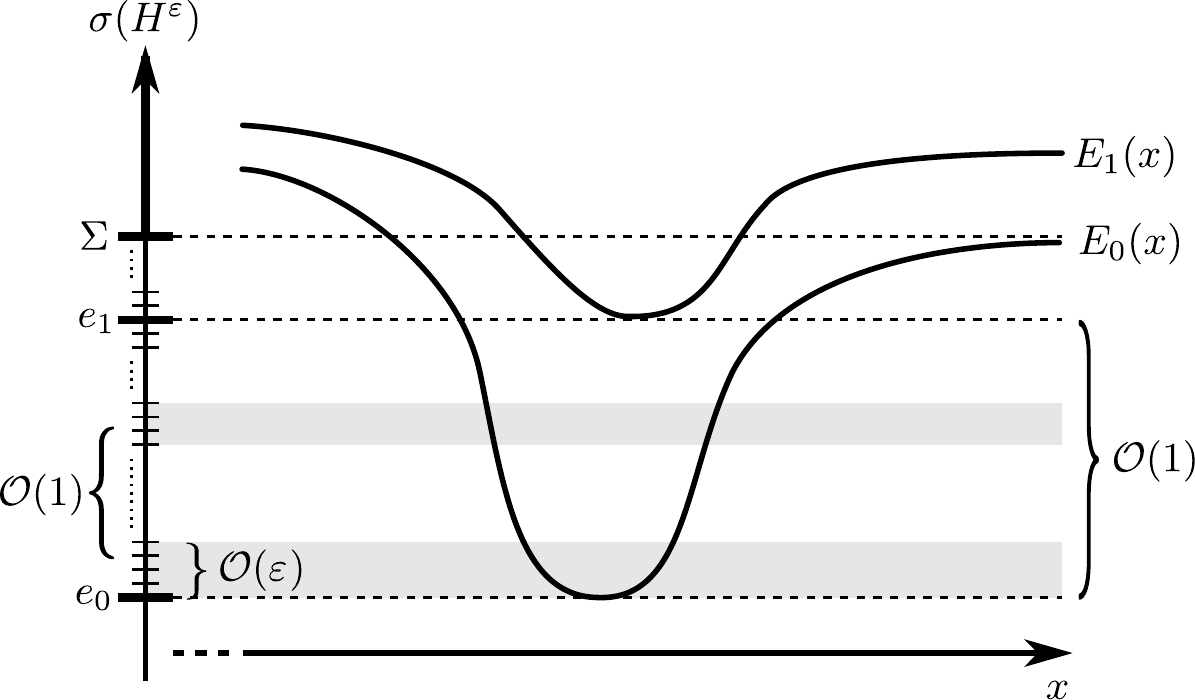}
 \caption{\label{fig2} The curves $E_0(x)$ and $E_1(x)$ are sketches of the lowest normal eigenvalues for a waveguide potential as depicted in Figure~\ref{fig1}(a). On the vertical axis the spectrum of $H^\veps$ is drawn: one expects spectral bands starting at the minima $e_0$ and~$e_1$ of the effective potentials with level spacing of order $\veps^2$. The continuum edge $\Sigma$ is determined by the threshold of $E_0$. Eigenstates in the lower shaded region vary on a $\sqrt{\veps}$-scale in the $x$-direction as indicated in Figure~\ref{fig1}(b). Eigenstates in the upper shaded region with energies of order one above $e_0$ have $\veps$-oscillations in  the $x$-direction as indicated in Figure~\ref{fig1}(c).
 } 
\end{figure}

Before we describe the general setup, it is instructive to first explain the scaling and the different energy scales  
 within the simple example  of a straight quantum wave guide in two dimensions. Let $x$ be the coordinate in the direction of propagation and $y$ the transversal direction. Saying that the potential $V(x,y)$ is (at least partially) confining in the $y$ direction just means, that the normal or transverse Hamiltonian
$H_{\rm n}(x) := -\Delta_y + V(x,y)$ has some eigenvalues $E_j(x)$ with localized eigenfunctions $\varphi_j(x,y)$, the constrained normal modes. 
For a sketch of such a potential see Figure~\ref{fig1}(a).
Now we would like to implement the
asymptotic limit of strong confinement in such a way, that the eigenfunctions of the scaled Hamiltonian $H^\veps_{\rm n}(x)$ become  localized on a length scale of order $\veps\ll 1$. This is done by scaling the potential $V^\veps (x,y) := V(x,y/\veps)$, which yields restoring forces of order $\veps^{-1}$. However, localization on a scale of order $\veps$  leads to kinetic energies of order $\veps^{-2}$. So in order to see localization one has to  increase not only the forces but also the potential energies  to the same level by putting 
\[
 H^\veps_{\rm n}(x) := - \Delta_y +  \veps^{-2} V(x,y/\veps)  \,.
 \]
 Then the normal energies and eigenfunctions are just $E^\veps_j(x) = \veps^{-2} E_j(x)$ and $\varphi_j^\veps(x,y) = \veps^{-\frac{1}{2}} \varphi_j(x,y/\veps)$.
 The full Hamiltonian becomes
 \[
 \tilde H^\veps = - \Delta_x - \Delta_y + \veps^{-2} V(x,y/\veps) \,.
 \]
In order to understand the asymptotic limit $\veps\to 0$   it is more convenient to rescale units of energy in such a way that the transverse energies are of order one again, i.e.\ to look at
 \begin{equation}\label{Heps1}
H^\veps :=  \veps^2  \tilde H^\veps = - \veps^2\Delta_x -\veps^2 \Delta_y +  V(x,y/\veps) \,.
 \end{equation}
Changing units of length in the transverse direction to $\tilde y = y/\veps$ finally leads to the form of the Hamiltonian
  \begin{equation}\label{HepsWG}
H_\veps   := \veps^2  \tilde H^\veps=  - \veps^2\Delta_x -  \Delta_{\tilde y} +  V(x,\tilde y) \,,
  \end{equation}
 for which the normal eigenfunctions $\varphi_j$ are independent of $\veps$ and the physical meaning of the asymptotic $\veps\to 0$  is most apparent. The limit of strong confinement really corresponds to the situation where the transversal modes are quantized with gaps of order one, while in the tangential direction the behavior is semiclassical and, in particular, the level spacing is of order $\veps^2$.
Here it is easy to guess the leading order effective Hamiltonian for the constrained system: on  the subspace of wave functions of the form $\Psi^\veps(x,\tilde y) =\varphi_j(x,\tilde y) \psi^\veps(x)$ the Hamiltonian  acts as
\begin{eqnarray}\label{Hexp}
H_\veps \Psi^\veps(x,\tilde y) &=& \left(- \veps^2\Delta_x  -  \Delta_{\tilde y} +  V(x,\tilde y) \right)  \varphi_j(x,\tilde y)\psi^\veps(x)\nonumber \\
 &=&
\varphi_j(x,\tilde y) \left[  \left(- \veps^2\Delta_x  + E_j(x) \right)  \psi^\veps(x) \right] \nonumber \\
 &&-\, 2( \veps  \nabla_x \varphi_j(x,\tilde y))(\veps\nabla_x \psi^\veps(x))\nonumber \\ &&-\, \veps^2\Delta_x \varphi_j(x,\tilde y)\psi^\veps(x) \,.
\end{eqnarray}
Defining the effective Hamiltonian by projecting back onto this subspace via $P(x) := |\varphi_j(x,\cdot)\rangle\langle \varphi_j(x,\cdot)|$ and integrating out $\tilde y$, one finds
\begin{eqnarray}\label{Heff}
H^\veps_{\rm eff} &=& ( - \I \veps\nabla_x - \I \veps \langle \varphi_j(x) | \nabla_x \varphi_j(x)\rangle )^2  + E_j(x)\\
&& + \,\veps^2 (\langle \nabla_x \varphi_j(x) | \nabla_x \varphi_j(x)\rangle-|\langle \varphi_j(x) | \nabla_x \varphi_j(x)\rangle|^2)\,.\nonumber
\end{eqnarray}
Here we see how the transversal eigenvalue $E_j(x)$ enters as an effective potential, the adiabatic potential, at leading order. E.g., the constraining potential sketched in Figure~\ref{fig1}(a) leads to an attractive effective potential sketched in Figure~\ref{fig2}.
The two energy scales referred to in the previous section now correspond to the following situations:  if one  assumes ``small'' tangential energies, i.e.\ 
$\langle \psi^\veps | - \veps^2 \Delta_x \psi^\veps\rangle = \O(\veps^2)$, then all terms in (\ref{Heff}) but the term involving the adiabatic potential $E_j(x)$ are of order~$\veps^2$. Thus the latter must be either constant or the kinetic energies will also become $\O(1)$ under the time evolution.
 Since it turns out that  the geometric potential in the case of non-straight wave guides is also of order~$\veps^2$, this explains why all authors interested in geometric effects up to now  assumed $E_j(x) \equiv$ const. 

 However, the natural scaling is to allow for tangential states $\psi^\veps$ with kinetic energies of order one, i.e.\ 
$\langle \psi^\veps | - \veps^2 \Delta_x \psi^\veps \rangle = \O(1)$. Then all energies in the system are of the same order and exchange of normal and tangential energies may occur. In particular, the tangential momentum operator $-\I\veps\nabla_x$ must be treated as being of order one  despite the factor $\veps$. This is the situation we will consider in the following.  

In the Figures~\ref{fig1} and \ref{fig2} we sketch the situation for a simple waveguide in a region where it widens slightly. Wave functions with tangential energies of order $\veps$ like in Figure~\ref{fig1}(b) yield the low lying part of the spectrum. General states with finite energy above the ground state, which include all states propagating through the wave guide, have tangential energies of order one and thus $\veps$-oscillations in $x$-direction, as indicated in Figure~\ref{fig1}(c). When the confining potential depends on $x$, there are, in general, no solutions with tangential kinetic energies of order $\veps^2$.
We also mention an extensive discussion of energy scales from a slightly different point of view in~\cite{BDT2}.

Before we explain the general model, it is instructive to mention two important points  on the level of this simple model. 
First of all one might want to add an ``external potential'' $W(x,y)$ which does not contribute to the confinement  and thus  is not scaled. We will allow for such an external potential with $W(x,y) = \mathcal{O}(1)$. Note that in the previous works focussing on geometry \cite{FrH,Mit} it was added on the small energy scale, i.e.\ $W(x,y) = \O(\veps^2)$.
The second remark is that with the energy scales there come also time scales. The time scale on which solutions with $\langle \psi^\veps | - \veps^2 \Delta_x \psi^\veps \rangle = \O(1)$ propagate distances of order one are times of order $\veps^{-1}$. This is because kinetic energies of order one for particles with  ``mass'' of order $\veps^{-2}$ yield  velocities of order $\veps$. The small energy solutions with $\langle \psi^\veps | - \veps^2 \Delta_x \psi^\veps \rangle = \O(\veps^2)$ propagate even slower, so here the natural time scale are times of order $\veps^{-2}$. The best results we can prove hold for even longer times, namely for times almost up to order~$\veps^{-3}$. Controlling the adiabatic decoupling for such long times makes the problem highly nontrivial. Roughly speaking, for times of order one the problem is just standard time-dependent perturbation theory. For times of order $\veps^{-1}$ one can use the ideas underlying the standard proof of the adiabatic theorem of quantum mechanics, see Subsection~\ref{SubSec1}. For longer times, however, one has to use ``super''-adiabatic, i.e.\ higher order adiabatic perturbation theory, see Subsection~\ref{subspace}.

 
\section{The adiabatic structure}
Here we first discuss in detail the model we consider and the assumptions involved. In the second subsection we introduce a horizontal momentum operator, the geometric generalization of $-\I\veps\nabla_x$ in the previous section,  which will play a crucial role in our results. Then we reveal the formal similarity with the Born-Oppenheimer approximation, before we explain the resulting adiabatic structure of the problem in the last subsection.


\subsection{Description of the model}\label{model}
Let $(\A,G)$ be a Riemannian manifold of dimension $d+k$ and $\C\subset \A$ a smooth submanifold of dimension $d$ without boundary and equipped with the induced metric $g=G|_\mathcal{C}$. We consider the Schr\"odinger equation on $\A$ with a potential $V_{\rm c}^\veps:\A\to \RRR$ that localizes all states from a certain subspace of $L^2(\A)$ close to $\C$ for small $\veps$, which will be made precise below.

So we want to start with fixed manifolds $\A$ and $\C$ and assume that the constraining potential $V_{\rm c}^\veps$ grows fast in the directions normal to $\C$ (strong restoring forces) while the variation along $\C$ is of order one (bounded tangential forces). 
This means we want to assume that
\begin{itemize}
\item \emph{normal derivatives of $V_{\rm c}^\veps$} are of order $\veps^{-1}$,
\item \emph{tangential derivatives of $V_{\rm c}^\veps$} are of order $1$,
\item \emph{all derivatives of the metric $G$} are of order $1$.
\end{itemize}
As explained in Subsection~\ref{SubSecScaling}, localization in the normal direction on a scale of order $\veps$  
 produces oscillations of order $\veps^{-1}$ in the tangential directions, too. 

When we introduce local coordinates $x=(x_i)_{i=1,\dots,d}$ in a neighborhood of $q\in\C$ and coordinates $N=(N_\alpha)_{\alpha=1,\dots,k}$ for the normal directions, the assumptions made above correspond, by the same reasoning that leads to (\ref{Heps1}) in Subsection~\ref{SubSecScaling},  to the Schr\"odinger equation
\begin{equation}\label{se2}
\I \,\partial_t \Psi^\veps \,=\, -\,\veps^2\Delta_{G} \Psi^\veps +V_{\rm c}(x, N/\veps )\Psi^\veps+W( y, N)\Psi^\veps
\end{equation}
where  $\Delta_{G}$ is the Laplace-Beltrami operator associated to $(\A,G)$ and the non-constraining potential $W$ may describe external forces.
Here the upper index $\veps$ at $\Psi$ means that we look at solutions with oscillations of order $\veps^{-1}$ and the $\veps^2$ in front of the kinetic energy ensures that these solutions have kinetic energies of order $1$.
For small $\veps$ at least some solutions of this equation concentrate close to the submanifold~$\mathcal{C}$. Therefore one expects that an effective Schr\"odinger equation on~$\mathcal{C}$ may be derived such that solutions $\psi_\veps(t)$ of the effective equation approximate the solutions $\Psi^\veps(t)$ of the full equation in a suitable way. 

\medskip

The scaling of the potential described in (\ref{se2})   depends on the choice of coordinates and cannot be implemented globally so naively. It just serves as a motivation for the following. In order to be able to implement a similar scaling globally we assume that the submanifold $\C$ has a tubular neighbourhood $\B$ of fixed diameter $\delta>0$. Within $\B$  it makes sense to speak of large derivatives  of $V^\veps$   with respect to the distance to $\C$. More precisely,  $\B$ can now be mapped to the $\delta$-neighbourhood $\B_\delta$ of the zero section in the normal bundle $N\C$. On $N\C$  the scaling of the potential as in (\ref{se2}) can be realized due   to its linear structure. Moreover,  for $\veps$ much smaller than $\delta$ all solutions below an arbitrary finite energy lie in $\B_{\delta/2}$ up to errors bounded by any power of $\veps$. Therefore it is possible to work completely on the normal bundle by constructing a diffeomorphism $\Phi: N\C\to \B$ and choosing a metric $\overline{g}$ on $N\C$ such that $\Phi$ is an isometry on~$\B_{\delta/2}$.

\medskip

To avoid all regularity problems we make the following assumption.

\medskip

\begin{assumption}\label{technisch}
The injectivity radii of $\A$ and $\C$ are strictly positive and all curvatures as well as their derivatives of arbitrary order are globally bounded. Furthermore, 
$V: N\mathcal{C}\to \RRR$ is smooth and bounded and arbitrary derivatives of $V$ are also globally bounded.
\end{assumption} 

In particular, this implies that $C$ and $N\C$ may be covered by coordinate neighborhoods such that for some $K\in\NNN$ not more than $K$ of them overlap at each point. This allows us to do all estimates in local coordinates.

\medskip

Our goal is now to find approximate solutions of the Schr\"odinger equation
\begin{equation*} 
\I \partial_t \Psi^\veps \,=\, - \veps^2 \Delta_{\overline{g}} \Psi^\veps + V_{\rm c}(q, N/\veps)\Psi^\veps +W( y, N)\Psi^\veps
\end{equation*}
on $\H=L^2(N\C,d\overline{\mu})$, where 
$\overline{g}$ is the pullback of $G$ via the diffeomorphism $\Phi$ on $\B_{\delta/2}$, suitably extended outside, and $d\overline{\mu}$ denotes the measure associated to $\overline{g}$. As expalined in the introduction it is helpful to rescale the normal coordinates to $n=N/\veps$. Then the equation reads
\begin{equation*} 
\I \partial_t \Psi^\veps = - \veps^2 (\Delta_{\overline{g}})^\veps \Psi^\veps + V_{\rm c}(q, n)\Psi^\veps +W( y, \veps n)\Psi^\veps =: H_\veps \Psi^\veps,
\end{equation*}
where $(\Delta_{\overline{g}})^\veps $ is the accordingly rescaled Laplacian,  whose expansion in $\veps$ is calculated in the appendices A2-A4. 


\subsection{The horizontal connection and the corresponding Laplacian}\label{horizontal}

Since we will think of the functions on $N\C$ as mappings from $\C$ to the functions on the fibers, the following objects will play a crucial role. Consider the bundle $\mathcal{E}_{\rm f}:=\{(q,\varphi)\,|\,q\in\C,\,\varphi\in L^2(N_q\C)\}$ over $\C$ which is obtained when the fibers $N_q\C$ of the normal bundle are replaced with $L^2(N_q\C)$ and the bundle structure of $N\C$ is lifted by lifting the action of ${\rm SO}(k)$ on the fibers to rotation of functions. We denote the set of all smooth sections of a hermitian bundle $\mathcal{E}$ by $\Gamma(\mathcal{E})$. 

\smallskip

For $\varphi\in\Gamma(\mathcal{E}_{\rm f})$ the \emph{horizontal connection} $\nabla^{\rm h}$
is defined by 
\begin{equation}
(\nabla^{\rm h}_\tau\varphi)(q,n)\;:=\;\frac{d}{ds}\Big|_{s=0}\varphi(w(s),v(s)),
\end{equation}  
where $\tau\in  T_q\C$ and $(w,v)\in C^1([-1,1],N\C)$ with
\begin{equation}\label{lift}
w(0)=q,\;\dot{w}(0)=\tau , \ \;\&\ \;v(0)=n,\;\nabla^\perp_{\dot{w}}v=0.
\end{equation}
Furthermore, $\Delta_{\rm h}$ is the bundle Laplacian associated to~$\nabla^{\rm h}$, i.e.\ defined by
\begin{eqnarray*}
\int_{\C}\int_{N_q\C}\psi^*\,\Delta_{\rm h}\psi\,dn\,d\mu 
&=& -\int_{\C}\int_{N_q\C}g^{ij}\,\overline{\nabla^{\rm h}_i\psi}\,\nabla^{\rm h}_j\psi\,dn\,d\mu
\end{eqnarray*}
where $g^{ij}$ is the inverse of the metric tensor~$g_{ij}$. 
Here and in the sequel we use the abstract index formalism including the convention that  one sums over repeated indices. Moreover, we will consistently use latin indices $i,j,..$ running from $1$ to $d$ for coordinates on $\C$, greek indices $\alpha,\beta,\dots$ running from $d+1$ to $d+k$ for the normal coordinates, and latin indices $a,b,..$ running from $1$ to $d+k$ for coordinates on the full normal bundle.

\smallskip

To obtain local expressions for these objects we fix $q\in\C$ and choose geodesic coordinate fields $\{\partial_{x_i}\}_{i=1,\dots,d}$ on an open neighborhood $\Omega$ of $q$ and an orthonormal trivializing frame $\{\nu_\alpha\}_{\alpha=1,\dots,k}$ of $N\Omega$. 
We define the connection coefficients $\omega^\gamma_{i\alpha}$ of the normal connection by $\nabla^\perp_{i}\nu_\alpha=\omega^\gamma_{i\alpha}\nu_\gamma$. 
Then the horizontal connection is given by 
\begin{equation}\label{horder}
\nabla^{\rm h}_{i}\varphi(q,n) \;=\; \partial_{i}\varphi(q,n)-\omega^\gamma_{i\alpha}n^\alpha\partial_{\gamma}\varphi(q,n),
\end{equation}
as was already shown by Mitchell \cite{Mit}, and it holds 
\begin{equation}\label{Laplace}
\Delta_{\rm h}\varphi\,=\,\mu^{-1}\big(\partial_{i}-\omega^\delta_{i\beta}n^\beta\partial_{\delta}\big)\mu \,g^{ij}\big(\partial_{j}-\omega^\gamma_{j\alpha}n^\alpha\partial_{\gamma}\big)\varphi
\end{equation}
with $\mu:=\det g_{ij}$.
The latter directly follows from the former and the definition of $\Delta_{\rm h}$. To obtain the former equation we note that for a normal vector field $v=n^\alpha\nu_\alpha$ over~$\C$ it holds \begin{equation}\label{Christoffel}
(\nabla^\perp_{i}v)^\gamma\;=\;\partial_{i}n^\gamma+\omega^\gamma_{i\alpha}n^\alpha.
\end{equation}
Now let $(w,v)\in C^1([-1,1],N\C)$ be as in (\ref{lift}).
Then by definition of $\nabla^{\rm h}$ we have
\begin{eqnarray*}
\nabla^{\rm h}_{i}\varphi(q,n) &=& \textstyle{\frac{d}{ds}}\big|_{s=0}\varphi(w(s),v(s))\\
&=& \textstyle{\frac{d}{ds}}\big|_{s=0}\varphi(w(s),n) \,+\, \textstyle{\frac{d}{ds}}\big|_{s=0}\varphi(q,v(s))\\
&=& \partial_{i}\varphi(q,n) \,+\, (\partial_{i}n^\gamma)\partial_{\gamma}\varphi(q,n)\\
&=& \partial_{i}\varphi(q,n) \,-\, \omega^\gamma_{i\alpha}n^\alpha\partial_{\gamma}\varphi(q,n)
\end{eqnarray*}
where we used (\ref{Christoffel}) and the choice of the curve $v$ in the last step.


\subsection{The splitting of the Laplace-Beltrami operator}

The basic idea for deriving an effective equation on the submanifold $\C$ is to split the Laplace-Beltrami operator on $N\C$ at leading order into a horizontal and a normal part relative to $\C$, similar to the splitting $-\Delta_x - \Delta_y$ in the simple example of Section~\ref{SubSecScaling}. To make this precise, 
first note that by construction at any point on  the zero section of $N\C$ (which we identify with $\C$ in the following) the tangent space splits into two orthogonal subspaces, one  tangent to $\C$ and one tangent to the fibre. Hence the metric tensor $\overline g$ and with it also the Laplace-Beltrami operator on $(N\C,\overline{g})$
splits into a sum
\[\Delta_{\overline{g}}\;=\; \Delta_g \,+\, \Delta_N,\]
where $\Delta_g$ is the Laplace-Beltrami operator on $\C$ and $\Delta_N$ is the euclidean Laplacian in the fibers $N_q\C \cong \RRR^k$ of the normal bundle. We note that $\Delta_g=\Delta_{\rm h}$ on functions that are constant on the fibers by (\ref{Laplace}). 
We will show that  also away from $\C$, i.e.\ globally on $N\C$, we can approximately split 
$\Delta_{\overline{g}}$   into a horizontal part, given by  $\Delta_{\rm h}$,  and the Laplacian in the fibre $\Delta_N$. The error grows linear with the distance $|N|$ to $\C$. 
Then the rescaling of the normal coordinates to $n=N/\veps$ yields that
\begin{equation}\label{aufspaltung} 
H_\veps \,=\, -\,\veps^2\Delta_{\rm h} -  \Delta_n + V_{\rm c}(q,n) + W(q,\veps n) +\O(\veps|n|) \,.
\end{equation} 
This operator has the same form as the Hamiltonian (\ref{BO}), which is the starting point for the time-dependent Born-Oppenheimer approximation, or the operator (\ref{HepsWG}) of our simple wave guide example. This suggests that   also in the general situation considered here adiabatic decoupling is the mechanism that yields effective Hamiltonians on $\C$. 

We now explain in more detail how to  achieve the above splitting of the Laplacian. 
An important step  is   to turn the measure on $N\C$ into product form. To do so we define
\begin{eqnarray*} 
M_\rho: L^2(N\C,d\overline{\mu}) &\to& L^2(N\C,dNd\mu)\,,\\
\Psi &\mapsto& M_\rho\,\Psi \,:=\, \rho^{-\frac{1}{2}}\,\Psi\,, 
\end{eqnarray*}
where $d N$ denotes   Lebesgue measure on the fibers $N_q\C\cong\RRR^k$ and $\rho = \frac{d\overline{\mu}}{d\mu\otimes d N}$ is the density of the original measure with respect to the product measure on $N\C$.
It is well-known that the unitary transformation of our Hamiltonian with $M_\rho$ leads to the occurence of  a purely geometric extra potential 
\[V_\rho=-\tfrac{1}{4}\overline{g}^{ab}(\partial_a \ln\rho)\,\partial_b \ln\rho+\tfrac{1}{2}\Delta_{\overline{g}}\ln\rho.\]
More precisely, it holds that
\begin{equation}\label{splitting}
M_\rho^* (-\Delta_{\overline{g}} )M_\rho \;=\; -\,\Delta_{\rm h}  \,-\,\Delta_N \,+\,V_\rho(q,N) \,+\, \O(|N|)\,,
\end{equation}
which is shown in the second to fourth appendix.
Therefore  after application of the unitary transformation $M_\rho$ and a Taylor expansion of $W$ the rescaled Hamiltonian $H_\veps$ is of the following form  close to $\C$:
\begin{eqnarray*}
H_\veps &=&  -\,\veps^2\Delta_{\rm h} \,-\,   \Delta_n \,+\, V_{\rm c}(q,n) \,+\,W(q,0) \\
&& \qquad\qquad\qquad\quad \,+\,\veps^2\, V_\rho(q,\veps n) \,+\, \O(\veps|n|)\,.
\end{eqnarray*}
We note that $-\,\veps^2\Delta_{\rm h}$ is of order $1$ on functions with oscillations of order $\veps^{-1}$. So the extra potential does not play a role for the leading order of the horizontal dynamics, unless the tangential kinetic energies are assumed to be small. Finally, it should be kept in mind that the remaining error term is small only when it is applied to functions that decay fast in the normal directions.


\subsection{Adiabatic decoupling}

Next we explain the principle of adiabatic decoupling in detail. For any $q\in\C$ we define the fiber Hamiltonian 
\[H_{\rm f} (q) \,=\, -\,\Delta_n + V_{\rm c}(q,n) +W(q,0)\]
on the Sobolev space $W^{2,2}(N_q{\cal C},dn)\subset L^2(N_q{\cal C},dn)$. We consider a $q$-dependent family of eigenvalues $E_{\rm f}(q)$ of multiplicity $M$, called an \emph{energy band} in the sequel, and an associated family of normalized eigenfunctions $\big(\varphi_{\rm f}^J(q)\big)_{J=1,\dots,M}$:
\begin{equation}\label{eigenwert} 
H_{\rm f} (q)\,\varphi_{\rm f}^J(q,\cdot) \;=\; E_{\rm f}(q) \,\varphi_{\rm f}^J(q,\cdot)\,.
\end{equation} 
By definition of $H_{\rm f}$ it holds
$H_\veps\;=\;H_{\rm f}\,-\,\veps^2\Delta_{\rm h}\,+\,\O(\veps)$ on states that decay fast enough.
Then states in
\[\mathcal{P}_0 \;:=\; \{ \varphi_{\rm f}^J(q,n)\psi_J(q)\,|\, \psi_J\in L^2(\mathcal{C},g)\}\subset L^2(N\mathcal{C})\]
are approximately invariant under the dynamics $\e^{-\I tH_\veps}$ for times of order~$1$. This is due to the fact that the associated projector $P_0$ defined by $P_0(q):=\big(|\varphi_{\rm f}^J\rangle\langle\varphi_{\rm f}^J |\big)(q)$ is a spectral projection of $H_{\rm f}$ and so we know that $[H_{\rm f},P_0]=0$, $[E_{\rm f},P_0]=0$, and $H_{\rm f}P_0=E_{\rm f}P_0$. Hence,
\begin{equation}\label{slowvariation}
[H_\veps,P_0]\;=\;[-\veps^2\Delta_{\rm h},P_0]\,+\,\O(\veps)\;=\;\O(\veps).
\end{equation}
More precisely, the solution of the full Schr\"odinger equation with initial value  $\Psi^\veps|_{t=0}=\psi^\veps_J\,\varphi_{\rm f}^J$ satisfies that
\[\Psi^\veps(t,q)\;=\; \varphi_{\rm f}^J(q,n)\psi^\veps_J(t,q) \,+\,\O(\veps |t|)\,,\]
where $\psi^\veps_J(t,q)$ solves the following effective Schr\"odinger equation on $\C$:
\begin{equation}\label{eff} 
\I\,\partial_t \psi^\veps_J(t,q) \;=\; -\,\veps^2  \Delta_{g} \psi^\veps_J(t,q) \,+\, E_{\rm f}(q)\,\psi^\veps_J(t,q).
\end{equation}
It is well-known that an equation of the form (\ref{eff}) does show interesting behavior only on the semiclassical  time scale $s=t/\veps$. The adiabatic principle, however, suggests that $\P_0$ may be expected to be invariant for such and even much longer times, if the energy band $(E_{\rm f})_q$ is separated by a gap  from the rest of the spectrum. Therefore we also assume the following.

\medskip

\begin{assumption}\label{luecke}
For all $q\in\C$ the fiber Hamiltonian $H_{\rm f}(q)$ has an eigenvalue $E_{\rm f}(q)$ of multiplicity $M$ such that
\[\inf_{q\in\C} \mathrm{dist}\big(E_{\rm f}(q), \,{\rm spec}(H_{\rm f}(q)\setminus E_{\rm f}(q))\,\big)\geq c_{\rm gap}>0\,.\]
In addition, there is a family of normalized eigenfunctions $\big(\varphi_{\rm f}^J(q)\big)_{J=1,\dots,M}$ which is globally smooth in $q$ (in particular, the corresponding eigenspace bundle is trivializable) and satisfies
\[\textstyle{\sup}_{q\in\C}\,\|{\rm e}^{\Lambda_0\langle n\rangle}\varphi_{\rm f}^J\|<\infty,\]
for $\langle n\rangle:=\sqrt{1+|n|^2}$, some $\Lambda_0>0$, and all $J$. \end{assumption}

An assumption about the decay is necessary because the error in the splitting $-\Delta_{\overline{g}}=-\veps^2\Delta_{\rm h}-\Delta_n+\O(\veps)$ is only small when applied to functions that decay fast enough, as was explained above. However, in lots of cases the decay is implied by the gap condition, in particular, for $E_{\rm f}$ below the continuous spectrum of $H_{\rm f}$. The assumption about triviality is necessary to get an effective equation on $L^2(\C,d\mu)\otimes\CCC^M$. If we dropped it, we would end up with an equation on a non-trivial rank-$M$ bundle over $L^2(\C,d\mu)$, which would complicate things quite a bit.


\section{Main results}\label{SectionMR}

Having revealed the adiabatic structure of the constraining Hamiltonian $H_\veps$ in the preceding section we have two different techniques at hand in order to deduce results about effective dynamics. 

On the one hand, it is possible to derive an analogue of the standard adiabatic theorem of quantum mechanics in order to show that the subspace $\P_0$ is invariant under $H_\veps$ for times of order $\veps^{-1}$ up to   errors of order $\veps$. This is analogous to the approach used in \cite{ST} in the context of the Born-Oppenheimer approximation and will be carried out in the first part of this section. It leads to the occurrence of a Berry connection that will be investigated in the second subsection.

In order to get a better approximation of the spectrum and/or to go to longer time scales for the dynamics, the usual adiabatic  technique relying on cancellation of errors due to  oscillations
is no longer practicable. However, the  general machinery of adiabatic perturbation theory, developed by Nenciu-Martinez-Sordoni in \cite{MS,NS,So} and Panati-Spohn-Teufel in \cite{PST1,PST2} and reviewed in \cite{Te},   allows to construct super-adiabatic subspaces that are invariant for times of order~$\veps^{-n_1}$ up to errors of order $\veps^{n_2}$ for arbitrary $n_1,n_2\in\NNN$. 
  More precisely, it allows to construct a projector $P_\veps$ which projects to a subspace $\P_\veps$ close to $\P_0$ and satisfies $[H_\veps,P_\veps]=\O(\veps^m)$ for any $m>1$. Adiabatic perturbation theory was  adapted to constrained quantum systems in~\cite{WT}. 
    For technical reasons it could be made rigorous only for $m\leq3$. The case $m=3$ seems enough for all applications though. Before we discuss the resulting effective Hamiltonian and the approximation of bound states in the last two subsections, we explain the construction of $P_\veps$ in the third subsection.


\subsection{Effective dynamics for times of order $\veps^{-1}$}\label{SubSec1}

The analogue of the adiabatic theorem, which provides effective dynamics for times of order $\veps^{-1}$, reads as follows:

\begin{theorem}\label{adiabatictheorem}
Fix $E_{\rm max}<\infty$ and denote by $\chi$ the characteristic function of $(-\infty,E_{\rm max}]$.
Let the energy band $E_{\rm f}$  and the family of normalized eigenfunctions $(\varphi_f^J)_{J=1,\dots,M}$ be as in Assumption \ref{luecke}. 

Define the operator $U_0:\P_0 \to L^2(\C,d\mu)\otimes\CCC^M$ by
\[U_0^{*}: L^2(\mathcal{C},d\mu)\otimes\CCC^M\to \P_0\,,\; \psi^\veps(q)\mapsto \varphi_{\rm f}^J(q,n)\,\psi^\veps_J(q)
\]
and $H_{\rm eff}^{(1)}:=U_0P_0H_\veps P_0U_0^*$. Then there is a $C<\infty$ such that for all $\veps$ small enough 
\begin{equation}\label{differenz}
\left\| \left( \e^{-\I H_\veps t} - U_0^*\e^{-\I H_{\rm eff}^{(1)} t } U_0\right)P_0\chi(H_\veps)\right\|<C\veps(1+ \veps|t|).
\end{equation}
Up to terms of order $\veps^2$ the first-order effective Hamiltonian $H_{\rm eff}^{(1)}$ is given by
\begin{eqnarray*}
\langle \psi^\veps|H_{\rm eff}^{(1)}\psi^\veps\rangle_{ \C } &=& \int_\mathcal{C} \left(g_{\rm eff}^{ijIJ}\,\overline{p_{{\rm eff}\,i}^{IK}\psi^\veps_K}\,p_{{\rm eff}\,j}^{JL}\psi^\veps_L + V_{\rm eff}^{IJ}\overline{\psi^\veps_I}\psi^\veps_J\right)d\mu
\end{eqnarray*}
with
\begin{eqnarray*}
p_{{\rm eff}\,j}^{JL} &=& -\I \veps \delta^{JL}\partial_j \,-\,\veps \,\langle \varphi_{\rm f}^J\,|\, \I  \nabla^{\rm h}_j\varphi_{\rm f}^L\rangle\\
g_{\rm eff}^{ijIJ} &=& g^{ij}\delta^{IJ} \,+\, \veps \,2{\rm II}^{ij}_\alpha \langle \varphi_{\rm f}^I\,|\, n^\alpha \varphi_{\rm f}^J\rangle\\
V_{\rm eff}^{IJ} &=& E_{\rm f}\delta^{IJ}+\veps (\partial_{\alpha}W)_{n=0}\langle \varphi_{\rm f}^I | n^\alpha  \varphi_{\rm f}^J\rangle,
\end{eqnarray*}
where ${\rm II}$ is the second fundamental form (see Appendix~1 for the definition), $\langle\,\cdot\,|\,\cdot\,\rangle_{ \C }$ is the scalar product on $L^2(\C,d\mu)$, and $\langle\,\cdot\,|\,\cdot\,\rangle$ is the scalar product on $L^2(\RRR^k,dN)$.
\end{theorem}

Via the operator $U_0^*$ it is, hence, possible to obtain approximate solutions of the original equation from the solutions of the effective equation. We point out that $P_0\chi(H_\veps)$ both cuts off high energies and produces initial states in $\P_0$. 
However, the cutoff energy $E_{\rm max}$ is arbitrary and, in particular, independent of $\veps$. It is only needed in order to get a uniform error bound, since for larger tangential energies the adiabatic decoupling becomes worse. Physically this is expected, since large tangential energies correspond to large tangential velocities and the separation of time-scales for the normal and the tangential motion,   which adiabatic decoupling is based on, breaks down for large velocities.

\smallskip

The effective Hamiltonian may be calculated using standard perturbation theory which is done in the last appendix. However, to verify that it yields effective dynamics on the relevant time scale $t=s/\veps$, i.e.\ to prove~(\ref{differenz}),  an additional adiabatic argument is needed. To make this clear we notice that the usual perturbative argument only yields an error of order $1$ for times of order $\veps^{-1}$:
Using that $U_0^*U_0=P_0$ and $U_0 U_0^*=1$ we have
\begin{eqnarray}\label{cook}
\lefteqn{\left(\e^{-\I H_\veps t}-U_0^*\e^{-\I U_0P_0H_\veps P_0U_0^*t}U_0\right)P_0}\\
&& \qquad=\,-\e^{-\I H_\veps t}\int_0^t\tfrac{d}{ds}\,\e^{\I H_\veps s}U_0^*\e^{-\I U_0 H_\veps U_0^*s}U_0\,ds\nonumber\\
&& \qquad=\,-\e^{-\I H_\veps t}\int_0^t \e^{\I H_\veps s}\I\left[H_\veps, P_0 \right]\,U_0^*\e^{-\I U_0 H_\veps U_0^*s}U_0\,ds\nonumber,
\end{eqnarray}
which is of order $\veps|t|$ by (\ref{slowvariation}) but cannot directly be seen to be small for times of order $\veps^{-1}$.
However,  adapting the calculation   in the derivation of the standard adiabatic theorem (see e.g.~\cite{Te}) shows that the integrand is, up to errors of order $\veps^2$, the time derivative of 
\[\e^{\I H_\veps s}\,\big(R_{H_{\rm f}}[H_\veps,P_0]-[P_0,H_\veps]R_{H_{\rm f}}\big)\,U_0^*\e^{-\I U_0 H_\veps U_0^*s}U_0,\] 
where $R_{H_{\rm f}}=P_0^\perp(H_{\rm f}-E_{\rm f})^{-1}P_0^\perp$ is the reduced resolvent. Therefore the time integral of this term yields an error of order $\veps$ independent of $t$, which shows that the whole error is, indeed, only of order $\veps(1+\veps|t|)$.

\smallskip

For times of order $\veps^{-1}$ the corrections of order $\veps$ yield relevant contributions: 

\begin{itemize}
\item The corrected momentum operator $p^\veps_{\rm eff}$ is a Berry connection on the $\CCC^M$-bundle over $\C$ where the effective wave function takes its values and therefore may give rise to topological and/or geometric phases (see the next subsection).
\item In general, all the corrections couple the effective internal degrees of freedom. If they, however, mutually commute, simultaneous diagonalization allows to split the effective $\CCC^M$-bundle into $M$ effective bundles of rank~$1$ locally. 
\item If the center of mass of $\overline{\varphi_{\rm f}^I}\varphi_{\rm f}^J(q)$ lies on $\C$ for all $q$, then $\langle \varphi_{\rm f}^I\,|\, n^\alpha \varphi_{\rm f}^J\rangle=0$. So both the correction to $g^\veps_{\rm eff}$ and to $V_{\rm eff}$ vanish in this case. In particular, a local splitting into $M$ effective bundles of rank~$1$ is possible in this case.
\end{itemize}


\subsection{The curvature of the Berry connection}\label{curvature}

In this  section we take a closer look at the induced Berry connection $p_{\rm eff}^{IJ} =  -\I \veps\partial_x\delta^{IJ}  -\veps \langle \varphi_{\rm f}^I | \I  \nabla^{\rm h}\varphi_{\rm f}^J\rangle$ that occurs in the effective Hamiltonian (see Theorem~\ref{adiabatictheorem}). 

\smallskip

For $M=1$, i.e.\ if the energy $E_{\rm f}$ is non-degenerate, it is simply a $U(1)$-connection that effects the dynamics similar to the vector potential of a magnetic field. If its curvature (the analogue of the magnetic field) is zero, one can achieve, at least locally, $p_{\rm eff} =  -\I \veps \partial_x$  by choosing a proper gauge, i.e.\ by choosing proper eigenfunctions~$\varphi_{\rm f}(x)$. But such a gauge might not exist globally and effects analogous to the Aharonov-Bohm effect may occur. In Section~\ref{TopEff} we give an example for a closed quantum wave guide without any external magnetic fields in which  such an effect occurs purely due to the geometry of the wave guide.

If the curvature of $p_{\rm eff}$ is non-zero, it does even locally change the dynamics at order $\veps$. In \cite{Ma}  Maraner identified the curvature as the origin of roto-vibrational couplings in simple molecular models. Moreover, further important effects are to be expected which are known for Berry connections from different areas: On the one hand, the anomalous velocity term in the semiclassical model for electrons in crystalline solids also stems from the curvature of a Berry connection, see \cite{SN, PST3}. On the other hand, in the Born-Oppenheimer approximation   the Berry connection term in the effective dynamics exactly cancels the effect of an external magnetic  field on the nuclei, see e.g.\ \cite{ST}. Neglecting the Berry term would lead to wrong physics: in the Born-Oppenheimer approximation a neutral molecule would suddenly react to the Lorentz force. 

\medskip

In the rest of this section we show how to obtain the following formula for the curvature of the Berry connection. 
To state it, we fix $q\in\C$ and choose again normal coordinate fields $\{\partial_{i}\}_{i=1,\dots,d}$ on an open neighborhood $\Omega$ of $q$. Then it holds  $[\partial_{i},\partial_{j}]=0$.  

\begin{proposition}\label{berryphase}
$\nabla^{\rm eff}:=\I\,p_{\rm eff}$ is a metric connection on the rank-$M$ bundle over $\C$ where the effective wave function takes its values. Its curvature vanishes for $M=1$ and otherwise is given by 
\begin{eqnarray*}
{\rm R}^{\nabla^{\rm eff}}_{IJij} &:=& \big(\nabla^{\rm eff}_{i} \nabla^{\rm eff}_{j} -\nabla^{\rm eff}_{j} \nabla^{\rm eff}_{i} -\nabla^{\rm eff}_{[\partial_i,\partial_j]}\big)_{IJ}\\
&=& \big(\nabla^{\rm eff}_{i} \nabla^{\rm eff}_{j} -\nabla^{\rm eff}_{j} \nabla^{\rm eff}_{i}\big)_{IJ}\\
&=& -\,\veps^2\,\langle\varphi_{\rm f}^I\,|\,{\rm R}^{\perp\,\gamma}_{\alpha ij}\,n^\alpha\,\partial_{\gamma}\varphi_{\rm f}^J\rangle(q)\\
&& +\,\veps^2\,\Big(\langle\nabla^{\rm h}_{i}\varphi_{\rm f}^I\,|\,\nabla^{\rm h}_{j}\varphi_{\rm f}^J\rangle-\langle\nabla^{\rm h}_{j}\varphi_{\rm f}^I\,|\,\nabla^{\rm h}_{i}\varphi_{\rm f}^J\rangle\Big)(q)\\
&& +\,\veps^2\,\Big(\langle\varphi_{\rm f}^I\,|\,\nabla^{\rm h}_{i}\varphi_{\rm f}^K\rangle\,\langle\varphi_{\rm f}^K\,|\,\nabla^{\rm h}_{j}\varphi_{\rm f}^J\rangle\\
&& \qquad\quad-\,\langle\varphi_{\rm f}^I\,|\,\nabla^{\rm h}_{j}\varphi_{\rm f}^K\rangle\,\langle\varphi_{\rm f}^K\,|\,\nabla^{\rm h}_{i}\varphi_{\rm f}^J\rangle\Big)(q)
\end{eqnarray*}
where ${\rm R}^\perp$ is the curvature of the normal connection (defined in the first appendix) and $\langle\,\cdot\,|\,\cdot\,\rangle$ is the scalar product on $L^2(\RRR^k,dN)$.
\end{proposition}

An analogue expression was derived by Mitchell in \cite{Mit} in the special case where $\varphi_{\rm f}^J$ is independent of~$q$ up to twisting. It was not realized that it always vanishes for $M=1$ though. 

\medskip

We will need that the connection $\nabla^{\rm h}$, which the normal connection induces on the bundle of functions over the normal fibers, is metric, i.e.\ $\partial_j\langle\varphi_1|\varphi_2\rangle_{\H_{\rm f}}=\langle\nabla^{\rm h}_j\varphi_1|\varphi_2\rangle_{\H_{\rm f}} +\langle\varphi_1|\nabla^{\rm h}_j\varphi_2\rangle_{\H_{\rm f}}$, and that its curvature is given by
\begin{equation}\label{curvature2}
{\rm R}^{\rm h}_{ij} \,:=\, \nabla^{\rm h}_{i} \nabla^{\rm h}_{j} -\nabla^{\rm h}_{j} \nabla^{\rm h}_{i} -\nabla^{\rm h}_{[\partial_i,\partial_j]}
\,=\, {\rm R}^{\perp\,\gamma}_{\alpha ij}\,n^\alpha\,\partial_{\gamma}.
\end{equation}

Since the normal connection is metric, its connection coefficients $\omega^\beta_{i\alpha}$ are anti-symmetric in $\alpha$ and $\beta$. So integration by parts yields
\begin{equation*}
\big\langle\omega^\gamma_{i\alpha}n^\alpha\partial_{\gamma}\varphi_1\big|\varphi_2\big\rangle(q)+\big\langle\varphi_1\big|\omega^\gamma_{i\alpha}n^\alpha\partial_{\gamma}\varphi_2\big\rangle(q)\;=\;0.
\end{equation*}
Therefore we have
\begin{eqnarray*}
\partial_j\langle\varphi_1|\varphi_2\rangle
&=& \langle\partial_{j}\varphi_1|\varphi_2\rangle\,+\,\langle\varphi_1|\partial_{j}\varphi_2\rangle\\
&=& \big\langle(\partial_{j}-\omega^\gamma_{i\alpha}n^\alpha\partial_{\gamma})\varphi_1\big|\varphi_2\big\rangle\\
&& \,+\,\big\langle\varphi_1\big|(\partial_{j}-\omega^\gamma_{i\alpha}n^\alpha\partial_{\gamma})\varphi_2\big\rangle\\
&\stackrel{(\ref{horder})}{=}& \langle\nabla^{\rm h}_j\varphi_1|\varphi_2\rangle+\langle\varphi_1|\nabla^{\rm h}_j\varphi_2\rangle.
\end{eqnarray*}

To compute the curvature of $\nabla^{\rm h}$ we notice that a simple calculation yields
\begin{eqnarray*}
{\rm R}^{\rm h}_{ij}
&=& \big(\partial_{i}\omega^\gamma_{j\alpha}-\partial_{j}\omega^\gamma_{i\alpha}\big)n^\alpha\partial_{\gamma}
+ \big[\omega^\delta_{i\alpha}n^\alpha\partial_{\delta},\omega^\gamma_{j\beta}n^\beta\partial_{\gamma}\big].
\end{eqnarray*}

Using the commutator identity
\begin{eqnarray*}
\big[\omega^\delta_{i\alpha}n^\alpha\partial_{\delta},\omega^\gamma_{j\beta}n^\beta\partial_{\gamma}\big]
&=& -\big(\omega^\beta_{i\alpha}\omega^\gamma_{j\beta}\,-\,\omega^\beta_{j\alpha}\omega^\gamma_{i\beta}\big)n^\alpha\partial_{\gamma}
\end{eqnarray*}
we obtain that
\begin{eqnarray*}
{\rm R}^{\rm h}_{ij}
&=& -\big(\partial_{x_i}\omega^\gamma_{j\alpha}-\partial_{x_j}\omega^\gamma_{i\alpha}+\omega^\beta_{i\alpha}\omega^\gamma_{j\beta}\,-\,\omega^\beta_{j\alpha}\omega^\gamma_{i\beta}\big)n^\alpha\partial_{n^\gamma}\\
&=& -R^{\perp\,\gamma}_{\alpha ij}n^\alpha\partial_\gamma,
\end{eqnarray*}
which was the claim. With this we can compute the curvature of the effective Berry connection. It is not difficult to verify that $\nabla^{\rm eff}$ is indeed a connection. Since $\nabla^{\rm h}$ is metric, we have that
\begin{eqnarray*}
\lefteqn{
\langle\varphi_{\rm f}^I|\nabla^{\rm h}_j\varphi_{\rm f}^J\rangle+\overline{\langle\varphi_{\rm f}^J|\nabla^{\rm h}_j\varphi_{\rm f}^I\rangle} }\\
&=& \langle\varphi_{\rm f}^I|\nabla^{\rm h}_j\varphi_{\rm f}^J\rangle +  \langle\nabla^{\rm h}_j\varphi_{\rm f}^I|\varphi_{\rm f}^J\rangle = \partial_j\langle\varphi_{\rm f}^I|\varphi_{\rm f}^J\rangle
=0.
\end{eqnarray*}
Thus the correction in $\nabla^{\rm eff}$ is anti-hermitean. Hence, for all $\psi_1,\psi_2:\C\to\CCC^M$
\begin{eqnarray*}
\veps\partial_j\big(\overline{\psi_1}\cdot\psi_2\big)
&=& \big(\overline{\veps\partial_j\psi_1}\big) \cdot\psi_2\,+\,\overline{\psi_1} \cdot\big(\veps\partial_j\psi_2\big)\\
&=& (\overline{\nabla^{\rm eff}_j\psi_1}) \cdot\psi_2\,+\,\overline{\psi_1} \cdot(\nabla^{\rm eff}_j\psi_2),
\end{eqnarray*}
which means that $\nabla^{\rm eff}$ is metric. Furthermore, this entails that the correction in $\nabla^{\rm eff}$ is purely imaginary for $M=1$. Since $\varphi_{\rm f}$ can be chosen real-valued for every $q\in\C$, which follows from $H_{\rm f}$ being real, we may gauge away the correction in an open neighborhood of any $q$. This implies that the curvature vanishes for $M=1$.

\smallskip

To compute the curvature of $\nabla^{\rm eff}$ for $M>1$ we calculate
\begin{eqnarray*}
{\rm R}^{\nabla^{\rm eff}}_{IJij}
&=& \veps^2\,\big(\nabla^{\rm eff}_{i} \nabla^{\rm eff}_{j} \,-\,\nabla^{\rm eff}_{j} \nabla^{\rm eff}_{i}\big)_{IJ} \\
&=& \veps^2\,\big(\partial_{i} \langle\varphi_{\rm f}^I |\nabla^{\rm h}_{j}\varphi_{\rm f}^J\rangle \,-\,\partial_{j} \langle\varphi_{\rm f}^I |\nabla^{\rm h}_{i}\varphi_{\rm f}^J\rangle\big) \\
&& +\,\veps^2\,\Big(\langle\varphi_{\rm f}^I\,|\,\nabla^{\rm h}_{i}\varphi_{\rm f}^K\rangle\,\langle\varphi_{\rm f}^K\,|\,\nabla^{\rm h}_{j}\varphi_{\rm f}^J\rangle\\
&& \qquad\quad-\,\langle\varphi_{\rm f}^I\,|\,\nabla^{\rm h}_{j}\varphi_{\rm f}^K\rangle\,\langle\varphi_{\rm f}^K\,|\,\nabla^{\rm h}_{i}\varphi_{\rm f}^J\rangle\Big).
\end{eqnarray*}
Using that $\nabla^{\rm h}$ is metric we obtain 
\begin{eqnarray*}
{\rm R}^{\nabla^{\rm eff}}_{IJij}
&=& \veps^2\,\langle\varphi_{\rm f}^I |{\rm R}^{\rm h}_{ij}\varphi_{\rm f}^J\rangle \\
&& +\veps^2\,\Big(\langle\nabla^{\rm h}_{i}\varphi_{\rm f}^I\,|\,\nabla^{\rm h}_{j}\varphi_{\rm f}^J\rangle-\langle\nabla^{\rm h}_{j}\varphi_{\rm f}^I\,|\,\nabla^{\rm h}_{i}\varphi_{\rm f}^J\rangle\Big)\\
&& +\veps^2\,\Big(\langle\varphi_{\rm f}^I\,|\,\nabla^{\rm h}_{i}\varphi_{\rm f}^K\rangle\,\langle\varphi_{\rm f}^K\,|\,\nabla^{\rm h}_{j}\varphi_{\rm f}^J\rangle\\
&& \qquad\quad-\langle\varphi_{\rm f}^I\,|\,\nabla^{\rm h}_{j}\varphi_{\rm f}^K\rangle\,\langle\varphi_{\rm f}^K\,|\,\nabla^{\rm h}_{i}\varphi_{\rm f}^J\rangle\Big).
\end{eqnarray*}
Then (\ref{curvature2}) yields  
\begin{eqnarray*}
{\rm R}^{\nabla^{\rm eff}}_{IJij} &=&  -\veps^2\,\langle\varphi_{\rm f}^I\,|\,{\rm R}^{\perp\,\gamma}_{\alpha ij}\,n^\alpha\,\partial_{\gamma}\varphi_{\rm f}^J\rangle\\
&& +\veps^2\,\Big(\langle\nabla^{\rm h}_{i}\varphi_{\rm f}^I\,|\,\nabla^{\rm h}_{j}\varphi_{\rm f}^J\rangle-\langle\nabla^{\rm h}_{j}\varphi_{\rm f}^I\,|\,\nabla^{\rm h}_{i}\varphi_{\rm f}^J\rangle\Big)\\
&& +\veps^2\,\Big(\langle\varphi_{\rm f}^I\,|\,\nabla^{\rm h}_{i}\varphi_{\rm f}^K\rangle\,\langle\varphi_{\rm f}^K\,|\,\nabla^{\rm h}_{j}\varphi_{\rm f}^J\rangle\\
&& \qquad\quad-\langle\varphi_{\rm f}^I\,|\,\nabla^{\rm h}_{j}\varphi_{\rm f}^K\rangle\,\langle\varphi_{\rm f}^K\,|\,\nabla^{\rm h}_{i}\varphi_{\rm f}^J\rangle\Big),
\end{eqnarray*}
which was to be shown.


\subsection{Construction of the superadiabatic subspace}\label{subspace}

There are several motivations and ways for further improving the result formulated in Theorem~\ref{adiabatictheorem}.  First of all one can aim at a better approximation, i.e.\ smaller error estimates. Next one can try to cover even longer time scales, i.e.\ times of order $\veps^{-2}$ and   beyond. These long time scales become relevant, e.g., when considering the propagation of states with tangential energies of order $\veps^2$ in wave guides where the energy band $E_{\rm f}$ is constant on all of $\C$, i.e.\ in the situation considered in earlier papers on geometric effects on constrained systems \cite{Ma1,Ma,Mit,FrH}. Last but not least one expects  
also the eigenvalues of the effective Hamiltonian  to be close to those of the full Hamiltonian and that one can recover, at least in a certain energy range, all eigenvalues of the full Hamiltonian in this way.

In order to achieve all three additional goals we show how to construct an effective Hamiltonian that is unitarily equivalent to the full Hamiltonian on a certain subspace of the full Hilbert space up to errors of order~$\veps^3$. To this end we use adiabatic perturbation theory\cite{PST1}. The strategy is to first associate a so called super-adiabatic subspace $\P_\veps$ with
 any  energy band  $E_{\rm f}$ satisfying Assumption~\ref{luecke}. The associated  projector $P_\veps$ turns out to be  uniquely fixed (up to terms of order $\veps^3$) by the requirement that it projects on $\P_0$ to leading order
and that the commutator $[H_\veps,P_\veps]$ is of order $\O(\veps^3)$. In a second step we construct a unitary $U_\veps$ mapping the range of $P_\veps$ to the Hilbert space of the constrained system $L^2(\C, d\mu)\otimes  \CCC^M$.

Then on the super-adiabatic subspace $H_\veps|_{\P_\veps} = P_\veps H_\veps P_\veps$ up to terms of order $\veps^3$. The effective Hamiltonian on $L^2(\C, d\mu)\otimes  \CCC^M$ is now given by  $H_{\rm eff}^{(2)} = U_\veps P_\veps H_\veps P_\veps U_\veps^*$ and solves all three problems mentioned above.

\medskip

We now explain this construction in detail. For the super-adiabatic projection we search for a bounded operator $P_\veps$ with
\begin{enumerate}
\item $P_\veps P_\veps\;=\;P_\veps$,
\item $P_\veps - P_0\; = \;\O(\veps)$,
\item $[H_\veps,P_\veps]\,\chi(H_\veps)\;=\;\O(\veps^3)$.
\end{enumerate}
Property i)  simply means that $P_\veps$ is an orthogonal projection, property ii) is the requirement to be close to the adiabatic projection $P_0$ and iii)
  says that $P_\veps\chi(H_\veps)\H$ is invariant under the Hamiltonian $H_\veps$ up to errors of order~$\veps^3$. 

\smallskip

Since we saw in (\ref{slowvariation}) that
$[H_\veps,P_0]=\O(\veps)$, it is consistent to make the   ansatz for 
  $P_\veps$ as an expansion in $\veps$ starting with $P_0$: 

\begin{eqnarray*}
P_\veps &=& P_0\,+\,\veps P_1\,+\,\veps^2 P_2\,+\,\O(\veps^3).
\end{eqnarray*}

We first construct $P_\veps$ in a formal way ignoring problems of boundedness. Afterwards we will explain how to obtain a well-defined projector and the associated unitary $U_\veps$.
We make the ansatz $P_1:=T_1^*P_0+P_0T_1$ with $T_1:\H\to\H$ to be determined. 
Using the expansion of $H_\veps=H_0+\veps H_1+\O(\veps^2)$ from the fourth appendix and assuming that $[P_1,-\veps^2\Delta_{\rm h}+E_{\rm f}]=\O(\veps)$ we have

\begin{eqnarray*}
[H_\veps,P_\veps]/\veps
&=& [H_0/\veps+H_1,P_0\,+\,\veps P_1]+\O(\veps)\\
&=& [H_0/\veps+H_1,P_0]\,+\,[H_0,P_1]+\O(\veps)\\
&=& [-\veps\Delta_{\rm h}+H_1,P_0]\,+\,[H_{\rm f}-E_{\rm f},P_1]+\O(\veps)\\
&=& (-\veps\Delta_{\rm h}+H_1)P_0-P_0(-\veps\Delta_{\rm h}+H_1)\\
&& +(H_{\rm f}-E_{\rm f})T_1^*P_0 -P_0T_1(H_{\rm f}-E_{\rm f})+\O(\veps)\,.
\end{eqnarray*}

We have to choose $T_1$ such that the first term is cancelled. Observing that the right hand side is off-diagonal with respect to $P_0$,
we may multiply with $P_0$ from the right and $P_0^\perp:=1-P_0$ from the left and vice versa to determine $P_1$. This leads to 

\begin{equation}\label{A1}
-\big(H_{\rm f}-E_{\rm f}\big)^{-1}P_0^\perp\big([-\veps\Delta_{\rm h},P_0]+H_1\big)\,P_0 \,=\, P_0^\perp T_1^*P_0
\end{equation}
and
\begin{equation}\label{A2}
-P_0\big([P_0,-\veps\Delta_{\rm h}]+H_1\big)P_0^\perp\big(H_{\rm f}-E_{\rm f}\big)^{-1}\,=\, P_0T_1P_0^\perp,
\end{equation}
where we have used that the operator $H_{\rm f}-E_{\rm f}$ is invertible on $P_0^\perp\H_{\rm f}$. In view of (\ref{A1}) and (\ref{A2}), we define $T_1$ by
\begin{eqnarray}\label{T1}
T_1 &:=& -P_0\big([P_0,-\veps\Delta_{\rm h}]+H_1\big)R_{H_{\rm f}}\nonumber\\
&& \quad+R_{H_{\rm f}}\big([-\veps\Delta_{\rm h},P_0]+H_1\big)P_0 
\end{eqnarray}
with $R_{H_{\rm f}}=P_0^\perp\big(H_{\rm f}-E_{\rm f}\big)^{-1}P_0^\perp$. $T_1$ is anti-symmetric so that $P^{(1)}:=P_0+\veps P_1=P_0+\veps (T_1^*P_0+ P_0T_1)$ automatically satisfies condition i) for $P_\veps$ up to first order: Due to $P_0^2=P_0$
\begin{eqnarray*}
P^{(1)}P^{(1)} &=& 
P^{(1)}+ \veps P_0(T_1^*+T_1)P_0+\O(\veps^2)\\
&=& P^{(1)}+\O(\veps^2).
\end{eqnarray*}
Moreover, it turns out that $P_1$ satisfies the assumption $[P_1,-\veps^2\Delta_{\rm h}+E_{\rm f}]=\O(\veps)$ made above, too.

In order to derive the form of the second order correction, we make the ansatz $P_2=T_1^*P_0T_1+T_2^*P_0+P_0T_2$ with some $T_2:\H\to\H$. The anti-symmetric part of $T_2$ is determined analogously with $T_1$ just by calculating the commutator $[P_\veps,H_\veps]$ up to second order and inverting $H_{\rm f}-E_{\rm f}$. One ends up with
\begin{eqnarray*}
(T_2-T_2^*)/2 &=& -P_0\big([P^{(1)},H^{(2)}]/\veps^2\big)R_{H_{\rm f}}\\
&& \,+R_{H_{\rm f}}\big([H^{(2)},P^{(1)}]/\veps^2\big)\,P_0
\end{eqnarray*}
with $H^{(2)}:=H_0+\veps H_1+\veps^2 H_2$. Note that $[H^{(2)},P^{(1)}]/\veps^2=\O(1)$ due to the construction of $P^{(1)}$.
The symmetric part is again determined by the first condition for $P_\veps$. Setting $P^{(2)}:=P^{(1)}\,+\,\veps^2 P_2$ we have
\begin{eqnarray*}
P^{(2)}P^{(2)} 
&=& P^{(2)}+\veps^2 P_0\big(T_1T_1^*+ T_2^*+T_2\big)P_0+\O(\veps^3),
\end{eqnarray*}
which forces $T_2^*+T_2=-T_1T_1^*$ in order to satisfy condition i) up to second order. 

\medskip

We note that $T_1$ is quadratic in the momentum (and $T_2$ even quartic) and will therefore not be bounded on the full Hilbert space and thus neither $P_\veps$. This is related to the well-known fact that for a quadratic dispersion relation adiabatic decoupling breaks down for momenta tending to infinity. The problem can be circumvented by cutting off high energies in the right place, which was carried out by Sordoni for the Born-Oppenheimer setting in \cite{So} and by Tenuta and Teufel for a model of non-relativistic QED in \cite{TT}. 

\smallskip

To do so we fix $E_{\rm max}<\infty$. Since $H_\veps$ is bounded from below, $E_-:=\inf\sigma(H_\veps)$ is finite. 
We choose $\tilde\chi\in C^\infty_0(\RRR,[0,1])$ with $\tilde\chi|_{(E_--1,E+1]}\equiv 1$ and ${\rm supp}\,\tilde\chi\subset(E_--2,E+2]$. Then we define
\begin{eqnarray}\label{preP}
P_\veps^{\tilde\chi}&:=& P_0 
+(P^{(2)}-P_0)\tilde\chi(H_\veps)\\
&& \,+\tilde\chi(H_\veps)(P^{(2)}-P_0)\big(1-\tilde\chi(H_\veps)\big)\nonumber
\end{eqnarray}
with $\tilde\chi(H_\veps)$ defined via the spectral theorem. We emphasize that $P_\veps^{\tilde\chi}$ is symmetric. 

It holds that $P_\veps^{\tilde\chi}-P_0=\O(\veps)$ in the sense of bounded operators. That is why for $\veps$ small enough a projector is obtained via the formula 
\begin{equation}\label{P}
P_\veps\;:=\;\frac{\I}{2\pi}\oint_{\Gamma}\big(P_\veps^{\tilde\chi}-z\big)^{-1}\,dz,
\end{equation}
where $\Gamma=\{z\in\CCC\,|\,|z-1|=1/2\}$ is the positively oriented circle around~$1$ (see e.g.\ \cite{DS}).
Denoting the associated subspace by $\P_\veps$ we define a unitary mapping $\tilde U_\veps:\P_\veps\to \P_0$ by the so-called Sz.-Nagy formula: 
\begin{equation}\label{Utilde}
\tilde U_\veps \,:=\,\big(P_0P_\veps+(1-P_0)(1-P_\veps)\big)\big(1-(P_\veps-P_0)^2\big)^{-\frac{1}{2}}.
\end{equation}

Then $U_\veps:=U_0\tilde U_\veps$ yields an isometry between $\P_\veps$ and $L^2(\C,d\mu)\otimes\CCC^M$. In \cite{WT} it is shown that $P_\veps$, indeed, satisfies i) to iii): 

\begin{proposition}\label{projector}
Fix $E_{\rm max}<\infty$. For all $\veps$ small enough $P_\veps$ is an orthogonal projection and $\tilde U_\veps$ is unitary. 
There are constants $C_i$ such that 
\begin{eqnarray}\label{invariance}
&\bullet& \|P_\veps-P_0\|_{\L(\H)}\;\leq\;C_1\,\veps\,,\nonumber\\
&\bullet& 
\|[H_\veps,P_\veps] \,\chi(H_\veps)\|_{\L(\H,\D(H_\veps))} \;\leq\;C_2\,\veps^3\nonumber\\
&\bullet& 
\|\langle n\rangle^l P_\veps\langle n\rangle^j\|_{\L(\H)}\;\leq\;C_3\quad\forall\ j,l\in\NNN_0,
\end{eqnarray}
with $\chi$ the characteristic function of $(-\infty,E]$.
\end{proposition}

The last estimate guarantees that the range of $P_\veps$ consists of states decaying faster than any polynomial, which is necessary to use the expansion of $H_\veps$ obtained in the fourth appendix.


\subsection{Effective dynamics for times of order $\veps^{-2}$} \label{SubSec2}

By combining the results of the previous section with standard perturbation theory we can conclude that for $H_{\rm eff}^{(2)}:=U_\veps P_\veps H_\veps P_\veps U_\veps^*$
we have
\[
\left\| \left( \e^{-\I H_\veps t/} - U_\veps^*\e^{-\I H_{\rm eff}^{(2)} t } U_\veps \right)P_\veps \chi(H_\veps)\right\|<C\veps^3|t|\,.
\]
In the super-adiabatic setting no further adiabatic averaging is needed. This clearly improves (\ref{differenz}) in the two ways anticipated: we get a better approximation and longer times. To get a simpler expression we can 
approximate $P_\veps$ and $U_\veps$ by $P_0$ and $U_0$  and find
\[
\left\| \left( \e^{-\I H_\veps t} - U_0^*\e^{-\I H_{\rm eff}^{(2)} t } U_0 \right)P_0 \chi(H_\veps)\right\|<C\veps(1 + \veps^2|t|)\,,
\]
i.e.\ still a good approximation for long times on the adiabatic subspace $\P_0$. However, we can not replace $H_{\rm eff}^{(2)}$ by $H_{\rm eff}^{(1)} =U_0 P_0 H_\veps P_0  U_0^*$ without loosing a factor $\veps$ in front of $|t|$ in the error. This is because the order $\veps^2$ terms in the effective Hamiltonian are relevant for times of order $\veps^{-2}$ and the expansion of the ``naive'' adiabatic Hamiltonian   $H_{\rm eff}^{(1)}$  yields incorrect second order terms. 

\medskip

So we still have to provide the correct second order expansion of the effective Hamiltonian $H_{\rm eff}^{(2)}$. Since the expression becomes quite complex and since we do not want to overburden the result, 
  we restrict ourselves to a non-degenerate energy band, i.e.\ with one-dimensional eigenspaces.

\begin{theorem}\label{calcHeff}
In addition to Assumptions \ref{technisch} and~\ref{luecke} assume that $E_{\rm f}$ is non-degenerate and that arbitrary derivatives of the corresponding family of eigenfunctions~$\varphi_{\rm f}$ are globally bounded.

Up to terms of order $\veps^3$ the second-order effective Hamiltonian $H^{(2)}_{{\rm eff}}$ is given by
\begin{eqnarray*}
\langle\psi^\veps|H^{(2)}_{{\rm eff}}\psi^\veps\rangle_\C
&=& \int_\mathcal{C} \left(g_{\rm eff}^{ij}\,\overline{p^{\rm eff}_{i}\psi^\veps}\,p^{\rm eff}_{j}\psi^\veps 
+V_{\rm eff}|\psi^\veps|^2\right.\\
&& \qquad\qquad\qquad  -\,\veps^2\,\overline{\psi^\veps}\,U_1^*R_{H_{\rm f}}U_1\,\psi^\veps\Big)d\mu,
\end{eqnarray*}
where
\begin{eqnarray*}
g_{\rm eff}^{ij} &=& g^{ij}+ \veps \,2{\rm II}^{ij}_\alpha \langle \varphi_{\rm f}| n^\alpha \varphi_{\rm f}\rangle +\veps^2\overline{\R}^{i\ j\;}_{\,\alpha\;\beta}\big\langle\varphi_{\rm f}\big|n^\alpha n^\beta\varphi_{\rm f}\big\rangle\\
&& \ +\veps^2\W_{\alpha l}^{i}g^{lm}\W_{\beta m}^{j}\big\langle\varphi_{\rm f}\big|3n^\alpha n^\beta \varphi_{\rm f}\big\rangle,\vspace{0.5cm}\\
p^{\rm eff}_{j} &=& -\I \veps \partial_j -\veps \langle \varphi_{\rm f}| \I  \nabla^{\rm h}_j\varphi_{\rm f}\rangle -\veps^2 \overline{\R}^{\ \ \gamma\;}_{j\alpha\;\beta}\langle\varphi_{\rm f}|{\textstyle \frac{2}{3}}n^\alpha n^\beta\I\partial_\gamma\varphi_{\rm f}\rangle\\
&& \,+\veps^2\W_\alpha^{ji}\big\langle\,\varphi_{\rm f}\,\big|\,2\,\big(n^\alpha-\langle\varphi_{\rm f}|n^\alpha\varphi_{\rm f}\rangle\big)
\I\nabla^{\rm h}_i\varphi_{\rm f}\,\big\rangle,\vspace{0.5cm}\\
V_{\rm eff} &=& E_{\rm f}+\veps (\partial_{\alpha}W)_{n=0}\langle \varphi_{\rm f} | n^\alpha  \varphi_{\rm f}\rangle\\
&& \qquad\qquad +\veps^2\,\big(V_{{\rm geom}}+V_{{\rm BH}}+V_{{\rm amb}}+W_2\big)\\
U_1 &=& 2g^{ij}\overline{\nabla^{\rm h}_i \varphi_{\rm f}}\partial_j+n^\alpha\varphi_{\rm f}\W^{ij}_\alpha\partial^2_{ij} -n^\alpha\varphi_{\rm f} (\partial_{\alpha}W)_{n=0}
\end{eqnarray*}
and
\begin{eqnarray*}
V_{{\rm geom}} &=& -{\textstyle \frac{1}{4}}\eta^{\alpha}\eta_\alpha+{\textstyle \frac{1}{2}}\R_{\;\;ij}^{ij}-{\textstyle \frac{1}{6}}\big(\overline{\R}_{\;\;ab}^{ab}+\overline{\R}_{\;\;aj}^{aj}+\overline{\R}_{\;\;ij}^{ij}\big),\\
V_{{\rm BH}} &=& g^{ij}\big\langle\nabla^{\rm h}_i\varphi_{\rm f}\big|\big(1-|\varphi_{\rm f}\rangle\langle\varphi_{\rm f}|\big)\nabla^{\rm h}_j\varphi_{\rm f}\big\rangle,\\
V_{{\rm amb}} &=& 
\overline{\R}^{\gamma\ \delta\;}_{\;\alpha\;\beta}\langle\partial_\gamma\varphi_{\rm f}|{\textstyle \frac{1}{3}}n^\alpha n^\beta\partial_\delta\varphi_{\rm f}\rangle,\\
W_2 &=&  (\partial^2_{\alpha\beta}W)_{n=0} \langle \varphi_{\rm f} | n^\alpha n^\beta \varphi_{\rm f}\rangle
\end{eqnarray*}
with $\W$ the Weingarten mapping, $\eta$ the mean curvature vector, $\R$ and $\overline{\R}$ the Riemann tensors of $\C$ and $\A$ (see Appendix 1 for the definitions).
\end{theorem}

This effective Hamiltonian is derived in \cite{WT}.  One might wonder whether the complicated form of the effective Hamiltonian renders the result useless for practical purposes. However, as explained in the introduction, the possibly much lower dimension of  $\C$ compared to that of $\A$
outweighs the more complicated form of the Hamiltonian. Moreover, the effective Hamiltonian is of a form that allows the use of semiclassical techniques  for a  further analysis. Finally, in practical applications typically only some of the terms appearing in the effective Hamiltonian are relevant. As an example we discuss the case of a quantum wave guide in Section~\ref{waveguides}.
At this point we only add some general remarks concerning  the numerous terms in $H^{(2)}_{\rm eff}$ and their consequences.

\begin{itemize}
\item The off-band coupling $U_1^*R_{H_{\rm f}}U_1$ can easily be checked to be gauge-invariant, i.e.\ not depending on the choice of $\varphi_{\rm f}$ but only on $P_0$. It occurs due to the replacement of $U_0$ by $U_\veps$ and thus is missed when one expands the naive adiabatic Hamiltonian $H_{\rm eff}^{(1)} =U_0 P_0 H_\veps P_0  U_0^*$. Even if one, in addition, uses standard perturbation theory in the fibers, still the first term in $U_1$, which originates from $[-\veps^2\Delta_{\rm h},P_0]$, would be missing.
\item Both $V_{{\rm BH}}$, an analogue of the so-called Born-Huang potential, and $V_{{\rm amb}}$, already found in \cite{Mit}, are also easily checked to be gauge-invariant, which justifies to call them extra potentials.
\item The occurence of the geometric potential $V_{{\rm geom}}$ has been stressed in the literature, in particular as the origin of curvature-induced bound states in quantum wave guides (reviewed by Duclos and Exner in~\cite{DE}), see Section \ref{waveguides} below.
\end{itemize}


\subsection{Approximation of bound states up to order $\veps^{3}$}\label{spectrum}

The unitary equivalence of $H_\veps$ and $H_{\rm eff}^{(2)}$ up to errors of order $\veps^3$ allows us to deduce that the lower parts of their spectra coincide up to errors of order $\veps^3$ when $E_{\rm f}$ is the ground state band.
The following result, which is   proved in \cite{WT}, shows how to obtain quasimodes of $H^\veps$ from the bound states of $H^{(2)}_{{\rm eff}}$ and vice versa.

\begin{theorem}\label{quasimodes}
Let $E_{\rm f}$ be a non-degenerate constraint energy band and let $U^\veps,H^{(2)}_{{\rm eff}}$ be the operators associated with $E_{\rm f}$ in the preceding subsection.

a) Let $E\in\RRR$. Then there is a $C<\infty$ such that for any family $(E_\veps)$ with $\limsup_{\veps\to0} E_\veps\,<\,E$ and all $\veps$ small enough  the following implications hold:
\begin{eqnarray*}
H^{(2)}_{{\rm eff}}\psi_\veps=E_\veps\psi_\veps &\ \Longrightarrow\ & \|( H_\veps-E_\veps) U_{\veps}^{*}\psi_\veps\| \leq  C\veps^3\|U_{\veps}^{*}\psi_\veps\|,\\ 
H_\veps\,\psi^\veps=E_\veps\psi^\veps &\ \Longrightarrow\ & \| (H^{(2)}_{{\rm eff}}-E_\veps)U_\veps\psi^\veps\| \leq C\veps^3 \|\psi^\veps\|.
\end{eqnarray*}

b) Let $E_{\rm f}(q)=\inf\sigma\big(H_{\rm f}(q)\big)$ for some (and thus for all) $q\in\C$ and define $E_1(q):=\inf\big(\sigma\big(H_{\rm f}(q)\big)\setminus E_{\rm f}(q)\big)$.
Let $(\psi^\veps)$ be a family with
\begin{equation}\label{threshold}
\limsup_{\veps\to0}\big\langle\psi^\veps\big|H_{\rm f}\psi^\veps\big\rangle<\inf_{q\in\C}E_1.
\end{equation}
Then there is $c>0$ such that $\|U_\veps\psi^\veps\|\geq c\,\|\psi^\veps\|$ for all $\veps$ small enough.
\end{theorem}

We recall that for any self-adjoint operator $H$ the bound $\|(H-\lambda)\psi\|<\delta\|\psi\|$ for $\lambda\in\RRR$ implies that $H$ has spectrum in the interval $[\lambda-\delta,\lambda+\delta]$. So a) i) entails that $H^\veps$ has an eigenvalue in an interval of length $2C\veps^3$ around~$E_\veps$, if one knows a priori that the spectrum of $H^\veps$ is discrete below the energy~$E$.
The statement b) ensures that a)~ii) really yields a quasimode for normal energies below $\inf_{q\in\C}E_1$, i.e.\ that 
\[
H_\veps\psi^\veps=E_\veps\psi^\veps \ \Longrightarrow\  \| (H^{(2)}_{{\rm eff}}-E_\veps)U_\veps\psi^\veps\|\leq {\textstyle\frac{C}{c}}\veps^3 \|U_\veps\psi^\veps\|\,.
\]

If the ambient manifold $\A$ is flat, then (\ref{threshold}) follows from
\begin{equation}\label{flatcase}
\limsup_{\veps\to0}\,\langle\psi^\veps|H_\veps\psi^\veps\rangle<\inf_{q\in\C}E_1-\sup_{(q,n)}(W_{n=0}-W)=:E_*.
\end{equation}
Therefore Theorem \ref{quasimodes}, in particular, implies that at least for flat $\A$ there is a one-to-one correspondence between the spectra of $H_\veps$ and $H^{(2)}_{{\rm eff}}$ below $E_*$.
In the example of Section~\ref{SubSecScaling} depicted in Figure~\ref{fig2} this implies that all eigenvalues of $H_\veps$ in the interval $[e_0,e_1)$ and the corresponding eigenfunctions are determined by the effective Hamiltonian of the ground state band $E_0$ modulo terms of order $\veps^3$.

\medskip

The bound states of $H^{(2)}_{{\rm eff}}$ can be approximated by the standard WKB construction. In the simplest case one obtains:

\begin{corollary}
Assume that $\A$ is flat and that $E_{\rm f}$ is a non-degenerate constraint energy band with $\inf E_{\rm f}<E_*$ and
$E_{\rm f}(q)=\inf\sigma\big(H_{\rm f}(q)\big)$ for all $q\in\C$. 
Let there be $q_0\in\C$ such that $E_{\rm f}(q_0)<E_{\rm f}(q)$ for all $q\neq q_0$ and $\big(\nabla^2_{i,j}E_{\rm f}\big)(q_0)$ is positive definite. 

Denote by $E_\ell(A)$ the $\ell$-th eigenvalue of a semi-bounded operator $A$, counted from the bottom of the spectrum. Then for any $\ell\in\NNN$
\begin{equation*}
E_\ell(H^\veps)\;=\;E_{\rm f}(q_0)\,+\,\veps E_\ell(H_{\rm HO})\,+\,\O(\veps^2),
\end{equation*}
where $H_{\rm HO}:=-\Delta_{\RRR^d}+\tfrac{1}{2}(\nabla^2_{\partial_{x^i},\partial_{x^j}}E_{\rm f})(q_0)x^ix^j$ is a harmonic oscillator on $\RRR^d$.
\end{corollary}


\section{Quantum wave guides}\label{waveguides}

In this section we look at the special case of a curve $\C$ in $\A=\RRR^3$ equipped with the euclidean metric. Such curves may model quantum wave guides which have been discussed theoretically for long times (see e.g.\ the review\cite{DE}) but are nowadays also investigated experimentally  (see e.g.\ the review~\cite{FZ}). 

In the first subsection we provide the expression for our effective Hamiltonian when applied to wave guides and make some general remarks about trapping and splitting of wave packets. In the second subsection we explain how to produce topological phases in closed wave guides. The effects on the spectrum of such wave guides are discussed in the last subsection.

\subsection{Trapping and splitting in quantum wave guides}\label{trapping}

We first look at infinite quantum wave guides. So let the curve $\C$ be given as a smooth injective $c:\RRR\to\RRR^3$ that is parametrized by arc length ($|\dot c|=1$). The mean curvature vector of $c$ is $\eta=\ddot c$ and its (exterior) curvature is $|\eta|$. Denoting by $\cdot$ the usual scalar product in $\RRR^3$ we define $y(n):=n\cdot\eta/|\eta|$ where $\eta\neq0$ and $y(n):=0$ elsewhere. By the Frenet formulas the Weingarten mapping satisfies $\W(\eta)=|\eta|^2$ (see e.g.\ \cite{DoC}) and $\W\equiv0$ on the orthogonal complement of $\eta$ (which is $N_q\C$ if $\eta(q)=0$). 

A normalized section of the tangent bundle $T\C$ is given by $\tau:=\dot c$. We extend this to an orthonormal frame of $T\C\times N\C$, where $N\C$ is the normal bundle, in the following way: We fix $q\in\C$, choose an arbitrary orthonormal basis of $N_q\C$, and take $\nu_1,\nu_2$ to be the parallel transport of this basis with respect to the normal connection $\nabla^\perp$ (defined in the first appendix) along the whole curve. This yields an orthonormal frame of~$N\C$. Together with $\tau$ we obtain an orthonormal frame of $T\C\times N\C$, which is sometimes called the Tang frame. We denote the coordinates with respect to $\tau$, $\nu_1$, and $\nu_2$ by $x$, $n_1$, and $n_2$ respectively. In these coordinates it holds $\nabla^{\rm h}=\partial_x$ (as can be seen from the coordinate formula (\ref{horder}) and the definition of the connection coefficients~$\omega$ in Appendix $1$).

\smallskip

Now let $E_{\rm f}$ and $(\varphi_{\rm f}^J)_J$ be as in Assumption \ref{luecke}. We start by spelling out the formula for $H_{\rm eff}^{(1)}$ from Theorem~\ref{adiabatictheorem}. Since $\C$ is one-dimensional and contractible, the families of $\varphi_{\rm f}^J$ can be chosen such that 
$p_{\rm eff}^\veps\equiv-\I\veps\partial_x$ globally. Then the first-order effective Hamiltonian is 
\begin{eqnarray}\label{fullwaveguide}
H_{\rm qwg}^{(1)} &=& -\I\veps\partial_x\big(1+\veps|\eta|\langle\varphi_{\rm f}^I |y\varphi_{\rm f}^J\rangle\big)\I\veps\partial_x+E_{\rm f}\nonumber\\
&& +\veps (\partial_{\alpha}W)_{n=0}\langle \varphi_{\rm f}^I | n^\alpha  \varphi_{\rm f}^J\rangle
\end{eqnarray} 
with $\langle\,\phi\,|\,\psi\,\rangle:=\int_{\RRR^2}\phi^*\,\psi\,dn_1dn_2$. 

\smallskip

For highly oscillating states $\psi$, i.e.\ with $\langle\psi|-\veps^2\partial^2_{xx}\psi\rangle\sim1$, the only term of order $1$ besides $-\veps^2\partial^2_{xx}$ is $E_{\rm f}$. So if $E_{\rm f}$ is constant, in particular, if the wave guide has constant cross section, the dynamics is free at leading order and, even more, the potential terms are of order~$\veps^2$. So they only become relevant for times of order~$\veps^{-2}$. However, a semiclassical wave packet $\psi$ covers distances of order~$\veps^{-1}$ on this time scale. Hence, for such~$\psi$ noteworthy trapping occurs only for very long wave guides!  

\smallskip

If we consider a straight wave guide, i.e.\ $\eta\equiv0$, the formula we end up with is the expected adiabatic approximation:
\begin{eqnarray*}\label{straightwaveguide}
H_{\rm qwg}^{(1)}|_{\eta\equiv0} &=& -\veps^2\partial^2_{xx} +E_{\rm f} +\veps (\partial_{\alpha}W)_{n=0}\langle \varphi_{\rm f}^I | n^\alpha  \varphi_{\rm f}^J\rangle.
\end{eqnarray*} 
We note that, although $\eta\equiv0$, the $x$-dependence of the constraining potential still allows us to model interesting situations. For example a beam splitter may be realized by fading a single-well into a double-well potential (see e.g.\ \cite{JS}).

\subsection{Topological phases in quantum wave circuits}\label{TopEff}

Up to now we have considered an infinite wave guide, which, of course, has the topology of a line. The only possible non-trivial topology for a curve $\C$ is that of a circle. We refer to a wave guide modeled over such a $\C$ as a \emph{quantum wave circuit}. In order to keep formulas simple and transparent,  we look at a so-called round circle, that is with constant $\eta$. Then the Tang frame from the preceding subsection is still globally smooth. However, because of the non-trivial topology our choices of the families $\varphi_{\rm f}^J$ made above are only possible locally but in general not globally. Therefore we rewrite (\ref{fullwaveguide}) without those choices. For the sake of brevity, we assume that $W$, the non-constraining part of the potential, is identically zero in the following. 
\begin{equation}\label{wavecircuit}
H_{\rm qwc}^{(1)} \;=\; p_{\rm eff}^*\big(1+\veps|\eta|\langle\varphi_{\rm f}^I|y\varphi_{\rm f}^J\rangle\big)p_{\rm eff}\,+\,E_{\rm f}
\end{equation} 
with $p_{\rm eff}=-\I\veps\partial_x+\veps\,\big\langle\varphi_{\rm f}^I\big|\I\partial_x\varphi_{\rm f}^J\big\rangle$. Although the curvature of the connection $\I p_{\rm eff}$ always vanishes, it may lead to a topological phase, which we will discuss next.

\medskip

Here and in the following subsection we again restrict ourselves to the case of a non-degenerate energy band~$E_{\rm f}$. We note that even for degenerate energy bands only abelian phases will occur because the fundamental group of the circle is generated by only one element. Let $x$ be a $2\pi$-periodic coordinate on the circle. The eigenfunction $\varphi_{\rm f}(x)$ associated to $E_{\rm f}$ can be chosen real-valued for each fixed $x$ because $H_{\rm f}$ is real. This associates a real line bundle to $E_{\rm f}$. From the topological point of view, there are exactly two real line bundles over the circle: the trivial one and the non-trivializable M\"obius band. In the former case the global section $\varphi_{\rm f}$ can be chosen real everywhere. This implies 
\[\langle\varphi_{\rm f}|\partial_x\varphi_{\rm f}\rangle=\tfrac{1}{2}\big(\langle\varphi_{\rm f}|\partial_x\varphi_{\rm f}\rangle+\langle\partial_x\varphi_{\rm f}|\varphi_{\rm f}\rangle\big)=\frac{\partial_x\langle\varphi_{\rm f}|\varphi_{\rm f}\rangle}{2}\equiv0,\] 
which results in $\I p_{\rm eff}=\veps\partial_x$. Thus there will be no topological phase in this case. We will now provide an example for the realization of the M\"obius band by a suitable constraining potential and show that, indeed, a topological phase occurs!

 \begin{figure}[h] 
\includegraphics[width=8  cm]{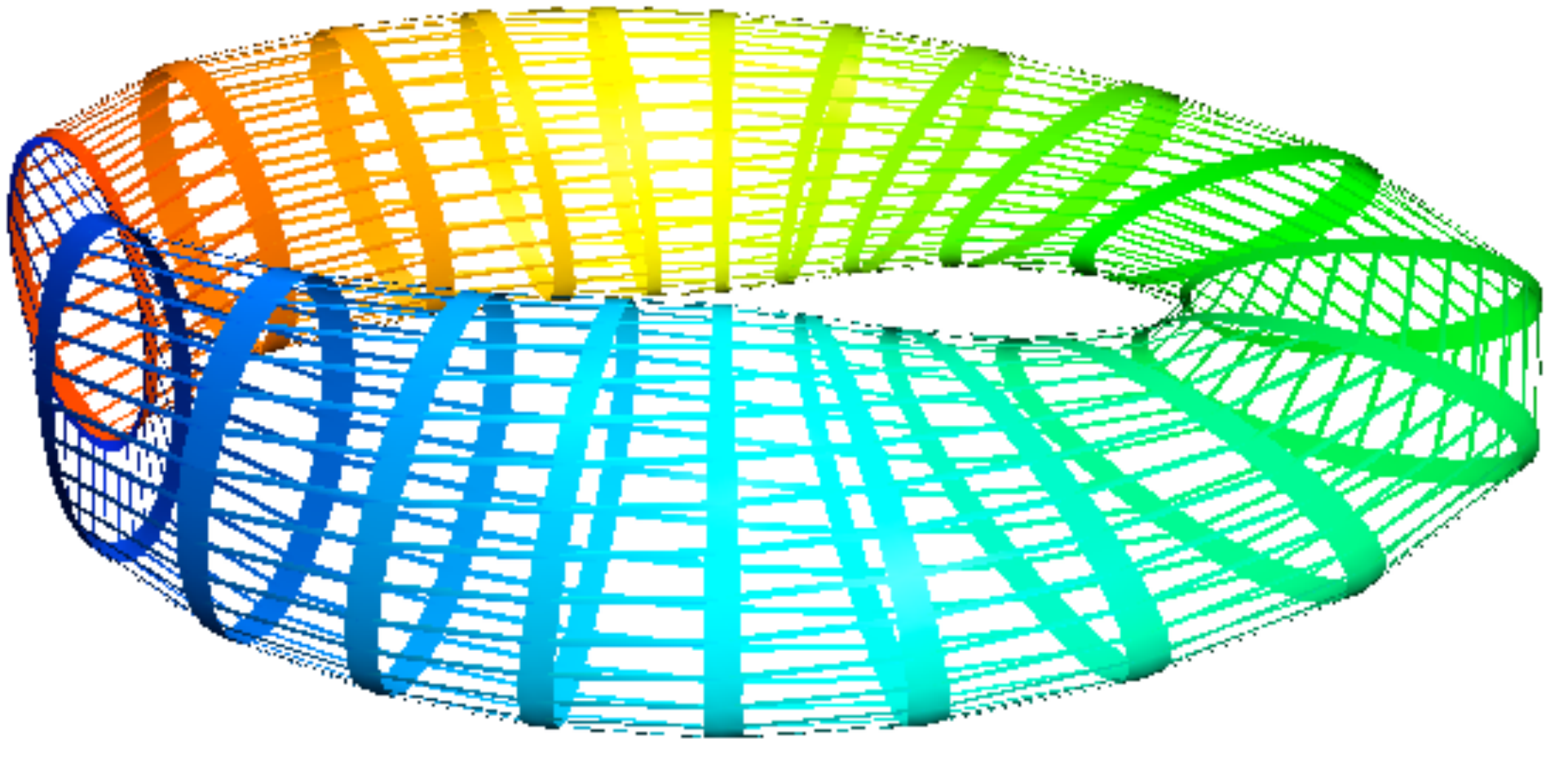}
 \caption{\label{fig4} Level set of a potential like $V_{\rm c}^{x/2}$ describing a ``M\"obius wave circuit''. 
} 
\end{figure}

\begin{figure}[h] 
\includegraphics[width=8.5 cm]{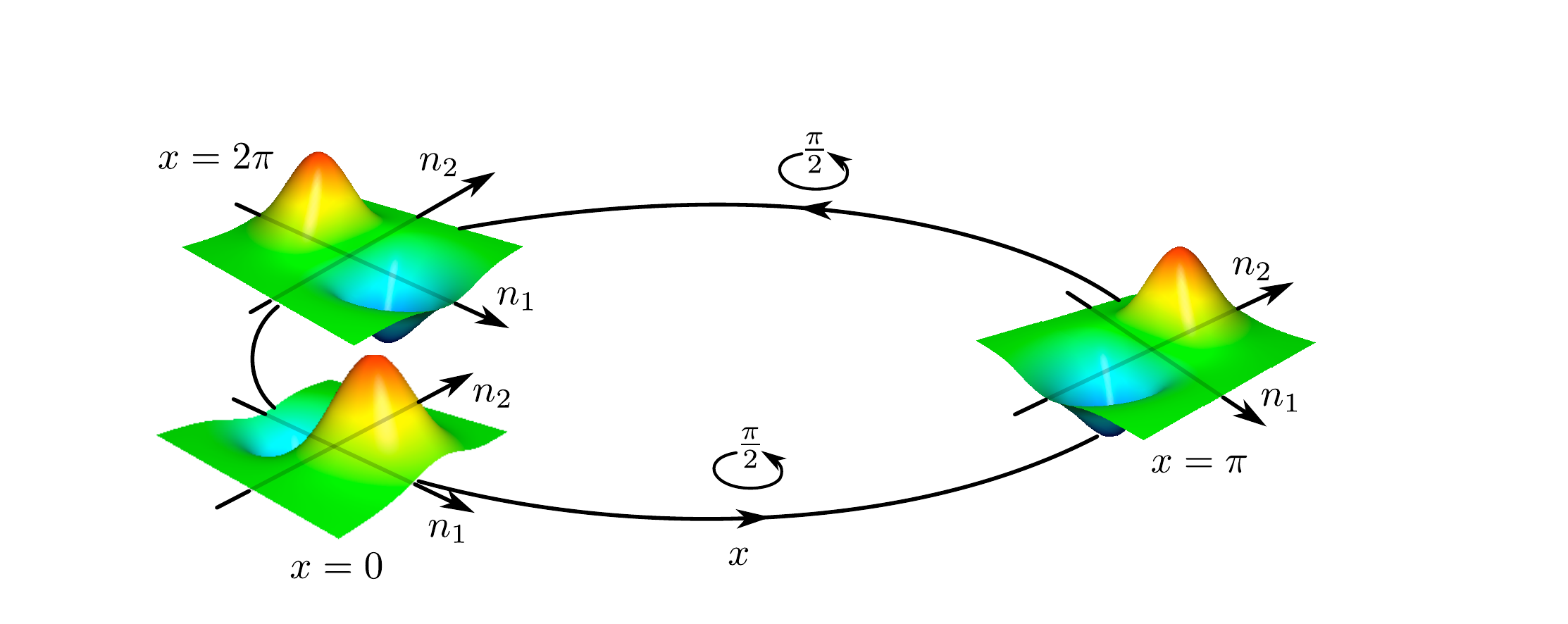}
 \caption{\label{fig3} First excited transverse eigenfunction in a ``M\"obius wave circuit'':  $n_1$ and $n_2$ are the normal elements of a global Tang frame and with respect to this frame the confining potential   $  V_{\rm c}^{x/2}$ twists by an angle of $\pi$ when going around the circuit once. Due to its symmetry $  V_{\rm c}^{x/2}$ is still globally smooth, see Figure~\ref{fig4}.
However, since the first excited state $\Phi_1$ has not the full symmetry of the potential, it changes sign after a twist by $\pi$. 
} 
\end{figure}

Let $\tilde V_{\rm c}\in C^\infty_{\rm b}(\RRR^2)$ have two orthogonal axes of reflection symmetry, i.e.\ in suitable coordinates 
\begin{equation}\label{symmetryV}
\tilde V_{\rm c}(-\tilde n_1,\tilde n_2)\;=\;\tilde V_{\rm c}(\tilde n_1,\tilde n_2)\;=\;\tilde V_{\rm c}(\tilde n_1,-\tilde n_2).
\end{equation} 
Then the real ground state $\Phi_0$  of $-\Delta_{\RRR^2}+\tilde V_{\rm c}$ with energy~$E_0$ is symmetric with respect to both reflections,
\[
\Phi_0(\tilde n_1,\tilde n_2)\;=\;\Phi_0(-\tilde n_1,\tilde n_2)\;=\;\Phi_0(\tilde n_1,-\tilde n_2)\,,
\]
while the first excited state $\Phi_1$, also taken real-valued, with energy $E_1$ is typically only symmetric with respect to one reflection and anti-symmetric with respect to the other one, e.g.\
\begin{equation}\label{symmetryPhi}
\Phi_1(\tilde n_1,\tilde n_2)\;=\;-\,\Phi_1(-\tilde n_1,\tilde n_2)\;=\;\Phi_1(\tilde n_1,-\tilde n_2)\,.
\end{equation} 
This is true in particular for a harmonic oscillator with different frequencies. 
As the potential constraining to the circle we let $\tilde V_{\rm c}$ perform half a twist along the circle, i.e.
\begin{eqnarray*}
\big(V_{\rm c}^{x/2}(x)\big)(n_1,n_2) &:=&
\tilde V_{\rm c}\big(\cos(x/2)n_1-\sin(x/2)n_2,\\
&& \qquad\quad \sin(x/2)n_1+\cos(x/2)n_2\big).
\end{eqnarray*} 
We note that due to (\ref{symmetryV}) this defines a smooth $V_{\rm c}^{x/2}$. Then
\begin{eqnarray*}
\big(\tilde\varphi_j(x)\big)(n_1,n_2)&:=& 
\Phi_j\big(\cos(x/2)n_1-\sin(x/2)n_2,\\
&& \qquad\quad \sin(x/2)n_1+\cos(x/2)n_2\big)
\end{eqnarray*} 
is an eigenfunction of $H_{\rm f}(x):=-\Delta_{\rm v}+V_{\rm c}(x)$ with eigenvalue $E_j$ for every $x$ and $j\in\{0,1\}$. However, while $\tilde \varphi_0$ is a smooth section of the corresponding eigenspace bundle,  $\tilde\varphi_1$ is not. For by (\ref{symmetryPhi}) it holds $\tilde\varphi_1(x)=-\tilde\varphi_1(x+2\pi)$ (see Figure \ref{fig3}). 
Still the complex eigenspace bundle admits a smooth non-vanishing section. A possible choice is $\varphi_1(x):=\e^{\I x/2}\tilde\varphi_1(x)$. Using (\ref{symmetryPhi})  we obtain that for the first excited band the effective Hamiltonian (\ref{wavecircuit}) reduces to
\begin{eqnarray*}
H_{\rm qwc,1} &=&  (-\I\veps\partial_x+\veps/2)^2 \,+\,E_1,
\end{eqnarray*} 
while for the ground state band it is
\begin{eqnarray*}
H_{\rm qwc,0} &=& -\veps^2\partial^2_{xx}\,+\,E_0.
\end{eqnarray*} 
This shows that depending on the symmetry of the normal eigenfunction
the twist by an angle of $\pi$ has different effects on the effective momentum operator in the effective Hamiltonian. 
With respect to the connection appearing in  $H_{\rm qwc,1}^\veps$ 
the holonomy of a closed loop $\gamma$ winding around the circle once is $h(\gamma)=\e^{\I\int_0^{2\pi}1/2\,dx}=-1$. Hence, the $1/2$ cannot be gauged away. Furthermore, a wave packet which travels around the circuit once accumulates a topological phase equal to $\pi$. 

\subsection{Effects of twisting and bending on the spectrum}

At second order in $\veps$ the effect of the topological phase in quantum wave circuits can also be seen in the level spacing of $H_{\rm qwc,j}$ and thus, with Theorem~\ref{quasimodes}, also in the spectrum of $H_\veps$. 
So we add the corrections of second order from Theorem \ref{calcHeff} to (\ref{wavecircuit}). Of course, all terms containing the inner curvature of $\C$ and $\A=\RRR^3$ vanish due to the flatness of $\C$ and $\A$ with the euclidean metric. 
\begin{eqnarray}\label{wavecircuit2}
H_{\rm qwc}^{(2)} &=& p_{\rm eff}^*\big(1+\veps|\eta|\langle\varphi_{\rm f}|y\varphi_{\rm f}\rangle+3\veps^2|\eta|^2\langle\varphi_{\rm f}|y^2\varphi_{\rm f}\rangle\big)p_{\rm eff}\nonumber\\
&& +E_{\rm f}+\veps^2\big(\langle\partial_x\varphi_{\rm f}|\partial_x\varphi_{\rm f}\rangle-|\langle\varphi_{\rm f}|\partial_x\varphi_{\rm f}\rangle|^2\big)\nonumber\\
&& -\veps^2\tfrac{|\eta|^2}{4}+\veps^2\Big(4\,\veps\partial_x\,\langle\partial_x\varphi_{\rm f}|R_{H_{\rm f}}\partial_x\varphi_{\rm f}\rangle\,\veps\partial_x\nonumber\\
&& \quad \,+\,4\,{\rm Re}\,|\eta|\,\veps\partial_{x}\,\langle \partial_x\varphi_{\rm f}|R_{H_{\rm f}}y\varphi_{\rm f}\rangle\,\veps^2\partial^2_{xx}\label{mixed}\nonumber\\
&& \quad\quad \,+\,|\eta|^2\,\veps^2\partial^2_{xx}\,\langle y\varphi_{\rm f}|R_{H_{\rm f}}y\varphi_{\rm f}\rangle\,\veps^2\partial^2_{xx}\Big)  
\end{eqnarray} 
with $p_{\rm eff}$ given by
\[-\I \veps \partial_x +\veps \langle \varphi_{\rm f}| \I  \partial_x\varphi_{\rm f}\rangle +\veps^2|\eta|\big\langle\varphi_{\rm f}\big|2\,\big(y-\langle\varphi_{\rm f}|y\varphi_{\rm f}\rangle\big)\partial_x\varphi_{\rm f}\big\rangle.\]

If $V_{\rm c}$ does not change its shape but only twists, $E_{\rm f}$ is constant and thus may be removed by redefining zero energy. 
Furthermore, since the remaining potential terms are of order $\veps^2$, the kinetic energy operator $-\veps^2\partial^2_{xx}$ will also be of order $\veps^2$ at the bottom of the spectrum. So $H_{\rm qwg}^\veps$ may be devided by $\veps^2$. Keeping only the leading order terms we arrive at
\begin{equation}\label{twist}
H_{{\rm qwc},j} = \tilde p_{\rm eff}^*\tilde p_{\rm eff}-\tfrac{|\eta|^2}{4}+\langle\partial_x\varphi_j|\partial_x\varphi_j\rangle-|\langle\varphi_j|\partial_x\varphi_j\rangle|^2
\end{equation} 
with $\tilde p_{\rm eff}:=-\I\partial_x+\big\langle\varphi_j\big|\I\partial_x\varphi_j\big\rangle$.
A simple calculation yields 
\begin{eqnarray*}
\lefteqn{\langle\partial_x\varphi_j|\partial_x\varphi_j\rangle-|\langle\varphi_j|\partial_x\varphi_j\rangle|^2}\\ 
&& =  \frac{1}{4}\int_{\RRR^2}|n_1\partial_{n_2}\Phi_j-n_2\partial_{n_1}\Phi_j|^2dn_1dn_2 \;=:\; L^2(\Phi_j)/4.
\end{eqnarray*}
We note that the integral is the expectation value of the squared angular momentum of $\Phi_{\rm f}$ and thus vanishes for a rotation-invariant $\Phi_{\rm f}$. So (\ref{twist}) shows that bending is attractive, while twisting is repulsive.

\smallskip

Since $|\eta|$ is constant, for $\ell\in \NNN_0$ the $\ell$-th eigenvalue of $H_{\rm qwc,1}$ is
\[
E_\ell(H_{\rm qwc,1}) = E_1 + \veps^2\left[(\ell+{\textstyle\frac{1}{2}})^2 + {\textstyle\frac{L^2(\Phi_1)-|\eta|^2}{4}}    \right]\,+\,\O(\veps^3),    
\]
while for $H_{\rm qwc,0}$ we find
\[
E_\ell(H_{\rm qwc,0}) = E_0 + \veps^2\left[\ell^2 + {\textstyle\frac{ L^2(\Phi_0)-|\eta|^2}{4}}  \right]\,+\,\O(\veps^3)   .
\]

We note that, although a constraining potential that twists along a circle was investigated by Maraner in detail in \cite{Ma1} and by Mitchell in \cite{Mit}, the effect discussed above was not found in both treatments. The reason for this is that they allowed only for whole rotations and not for half ones to avoid the non-smoothness of $\tilde\varphi_1$.  

\medskip

There is a wide literature on the spectrum of a quantum wave guide which is arbitrarily bent and twisted (see the review \cite{K} by Krej$\check{{\rm c}}$i$\check{\rm r}$\'ik). In general, the twisting assumption means that there is $\theta\in C^\infty_{\rm b}(\RRR)$ and $\tilde V_{\rm c}\in C^\infty_{\rm b}(\RRR^2)$ such that the constraining potential has the form:
\begin{eqnarray*}
\big(V_{\rm c}^\theta(x)\big)(n_1,n_2)&:=& 
\tilde V_{\rm c}\big(n_1\cos\theta(x)-n_2\sin\theta(x),\\
&& \qquad\qquad n_1\sin\theta(x)+n_2\cos\theta(x)\big).
\end{eqnarray*} 
Then the family of eigenfunctions $\varphi_{\rm f}$ may be chosen as
\begin{eqnarray*}
\big(\varphi_{\rm f}(x)\big)(n_1,n_2)&:=& 
\Phi_{\rm f}\big(n_1\cos\theta(x)-n_2\sin\theta(x),\\
&& \qquad\qquad n_1\sin\theta(x)+n_2\cos\theta(x)\big)
\end{eqnarray*} 
for an eigenfunction $\Phi_{\rm f}$ of $-\Delta_{\RRR^2}+\tilde V_{\rm c}(x)$ with eigenvalue~$E_{\rm f}$. It is easy to generalize the discussion above to a wave circuit whose curvature and potential twist are non-constant. Then the $\ell$-th eigenvalue of $H_\veps$ is given by
\begin{equation*}
E_\ell(H_\veps)\,=\,E_{\rm f}+\veps^2 E_\ell(H_{\rm twist}^\theta)+\O(\veps^3),
\end{equation*}
where $E_\ell(H_{\rm twist}^\theta)$ is the  $\ell$-th eigenvalue of the following operator:
\begin{equation*}
H_{\rm twist}^\theta\,:=\,\tilde p_{\rm eff}^*\tilde p_{\rm eff}-|\eta|^2/4+L^2(\Phi_{\rm f})\,\dot\theta^2/4.
\end{equation*}
with $\tilde p_{\rm eff}:=-\I\partial_x+\big\langle\varphi_{\rm f}\big|\I\big(\partial_x+(\nu_\alpha\cdot\partial_x\nu_\beta)n^\alpha\partial_\beta\big)\varphi_{\rm f}\big\rangle$. This generalizes results by Bouchitt\'e, Mascarenhas and Trabucho~\cite{BMT} and by Borisov and Cardone~\cite{BC} for wave guides to wave circuits. We note that for the Tang frame $\nu_\alpha\cdot\partial_x\nu_\beta\equiv0$ but for an arbitrarily curved wave circuit the Tang frame is, in general, not globally smooth anymore. Anyway the normal bundle is still trivializable because it inherits the orientation of $\RRR^3$ and every orientable vector bundle over a curve is trivializable.


\section{Conclusions}

While all earlier results on constrained quantum systems had to focus either on a certain energy regime or on special geometries, we have presented here results, both on    the dynamics and on the spectrum,  that cover all relevant energy regimes in general geometries  (recall Figure~\ref{fig2}). 

We point out that our results on dynamics (Theorem \ref{adiabatictheorem} and Theorem \ref{calcHeff}) are true for all bound state and scattering energies, as long as oscillations faster than $\veps^{-1}$ are excluded. The same is true for the quasimodes of the full Hamiltonian $H^\veps$ constructed from those of the effective Hamiltonian (Theorem~\ref{quasimodes}). 

\smallskip

Furthermore, we have applied our results to quantum wave guides and obtained for the first time the complete second order effective Hamiltonian (\ref{wavecircuit2}). In contrast to earlier theoretical results it applies also to wave circuits, i.e.\ closed wave guides. Here the effect of an abelian topological phase is observable both in the spectrum and in the dynamics. We believe that as a next step it would be interesting to apply our results to simple examples from molecular dynamics, like those that were treated for small kinetic energies by Maraner in \cite{Ma}. Here also the curvature of the effective Berry connection, calculated in Proposition \ref{berryphase}, should play a role. Note that it did not show up for quantum wave guides because of the one-dimensional constraint manifold. 


\section*{Appendix}


\subsection*{A1\ Geometry of submanifolds}

We recall here some standard concepts from Riemannian geometry. 
For further information see e.g.\ \cite{L}. As before we use the abstract index formalism including the convention that  one sums over repeated indices. Moreover, we will consistently use latin indices $a,b,..$ running from $1$ to $d+k$ for coordinates on a general manifold, latin indices $i,j,..$ running from $1$ to $d$ for coordinates on a submanifold,  and greek indices $\alpha,\beta,\dots$ running from $d+1$ to $d+k$ for coordinates in the normal spaces of a submanifold.

\smallskip

First we give the definition of the Riemann tensor we use (the order of the indices varies in the literature!). 

\begin{definition}
Let $(\M,g)$ be a Riemannian manifold with Levi-Civita connection $\nabla$. 
Let $(\tau_a)_{a}$ be a set of local coordinate vector fields.

\smallskip

i) The \emph{Christoffel symbols} $\Gamma$ of $\nabla$ are defined by 
\[\nabla_{a}\tau_b\;=\;\Gamma_{ab}^c\tau_c.\]

ii) The \emph{Riemann tensor} $\R$ is given by
\[\R^a_{\;bcd}\;:=\;\big(\nabla_{c}\nabla_{d}\tau_b-\nabla_{d}\nabla_{c}\tau_b
\big)^a\,.\]
\end{definition}

As usual by raising and lowering indices we mean to shift covariant to contravariant coordinates and vice versa. 

\medskip

Now we turn to the basic objects related to the exterior curvature of a submanifold of arbitrary codimension.

\begin{definition}
Let $(\A,\overline{g})$ be a Riemannian manifold with Levi-Civita connection $\overline{\nabla}$. 
Let $\C\subset\A$ be a submanifold equipped with the induced metric $g=\overline{g}|_{\C}$. Denote by $N\C$ the normal bundle of $\C$. 
Let $(\tau_i)_{i}$ be a set of local coordinate vector fields of $\C$ and $(\nu_\alpha)_\alpha$ a local orthonormal frame of $N\C$. 

\smallskip
 
i) The \emph{Weingarten mapping} $\W$ is given by 
\[\W_{\alpha j}^{i}\;:=\;\big(-\overline{\nabla}_j \nu_\alpha\big)^i.\]
 
ii) The \emph{second fundamental form} ${\rm II}$ is defined by
\[{\rm II}^{\alpha}_{ij}\;:=\;\big(\overline{\nabla}_j \tau_i\big)^\alpha.\]

iii) The \emph{mean curvature normal} $\eta$ is defined by
\[\eta_\alpha\;=\;\W_{\alpha j}^{j}.\]
\end{definition}

The relations and symmetry properties of $\W$ and ${\rm II}$ for hypersurfaces also hold when the codimension is greater than one:
\begin{eqnarray}\label{symmetry}
{\rm II}_{\alpha j}^{i} \;=\; \W_{\alpha j}^{i}
&=& \W_{\alpha i}^{j} \;=\; {\rm II}_{\alpha i}^{j}.
\end{eqnarray}

\smallskip

Finally, we provide the definitions of the objects that characterize the geometry of the normal bundle.

\begin{definition}
i) We define the \emph{normal connection} $\nabla^\perp$ to be the bundle connection on the normal bundle given via
\begin{equation*}
(\nabla^\perp_j\nu_\alpha)^a\;=\;(\overline{\nabla}_j\nu_\alpha)^a,
\end{equation*}
for $a=d+1,\dots,d+k$ and $(\nabla^\perp_j\nu_\alpha)^a=0$ for $a=1,\dots,d$.

ii) The \emph{connection coefficients} $\omega$ of $\nabla^\perp$ are defined by 
\[\nabla^\perp_i\nu_\alpha\;=\;\omega_{i\alpha}^\beta\nu_\beta.\]

iii) The \emph{normal curvature tensor} ${\rm R}^\perp$ is defined by
\[\R^{\perp\,\alpha}_{\;\beta ij}\;:=\;\big(\nabla^\perp_{i}\nabla^\perp_{j}\nu_\beta-\nabla^\perp_{j}\nabla^\perp_{i}\nu_\beta\big)^\alpha.\]
\end{definition}

Due to the anti-symmetries of any curvature tensor the normal curvature tensor ${\rm R}^\perp$ is identically zero, when the dimension or the codimension of $\C$ is smaller than $2$. 

\smallskip

When we set $\W_{\alpha j}^{\;\,a}=0$ for $a=d+1,\dots,d+k$, the Weingarten equation
\begin{equation}\label{Weingarten}
(\nabla^\perp_j\nu_\alpha)^a\;=\;(\overline{\nabla}_j\nu_\alpha)^a+\W_{\alpha j}^{\;\,a}
\end{equation} 
 is a direct consequence of the definitions.


\subsection*{A2\ Expansion of the metric tensor}

In order to expand the Hamiltonian $H_\veps$ in powers of $\veps$ it is crucial to expand the metric  $\overline{g}$ on the normal bundle~$N\C$ around $\C$ because the Laplace-Beltrami operator depends on it. Such expansions were carried out in almost any work on constrained quantum systems, however, in various generalities and up to varying orders. Here we provide a simple derivation for an arbitrary submanifold of a curved ambient manifold but only up to first order. However, this is enough in order to obtain Theorem \ref{adiabatictheorem}.

\smallskip

Fix $q\in\C$. Let $(\tau_i)_{i}$ be a set of local coordinate vector fields of $\C$ and $(\nu_\alpha)_\alpha$ a local orthonormal frame of $N\C$. Furthermore, let $\phi(q):=\big(\phi^1(q),\dots,\phi^{d+k}(q)\big)$ be the local expression for the isometric embedding of $\C$ into $N\C$ and $\Phi(q,N)$ be the exponential map in each fiber. Then, by definition of the exponential map, $\Phi(q,N)=\gamma(1)$ where $\gamma(s)$ is the geodesic on $N\C$ starting at $\phi(q)$ with $\dot\gamma(0)=N^\alpha\nu_\alpha$. 

Let $(y^a)_{a=1,\dots,d+k}$ be Riemannian normal coordinates on $(N\C,\overline{g})$ around $\phi(q)$ and $\Gamma^a_{bc}$ the  associated Christoffel symbols of the Levi-Civita connection. Due to the geodesic equation
$\ddot\gamma^a(s)=-\Gamma^a_{bc}\big(\gamma(s)\big)\dot\gamma^b(s)\dot\gamma^c(s)$
a Taylor expansion around $s=0$ yields
\begin{eqnarray*}
\gamma^a(s) &=& \phi^a(q)+sN^\alpha\nu_\alpha^a(q)\\
&& -\tfrac{s^2}{2}N^\alpha N^\beta\Gamma^a_{bc}\big(\phi(q)\big)\nu_\alpha^b(q).\nu_\beta^c(q)+\O(s^3)
\end{eqnarray*}
Evaluating at $s=1$ we obtain that
\begin{eqnarray*}
\Phi^a(q,N) &=& \phi^a(q)+N^\alpha\nu_\alpha^a(q)\\ 
&&-\tfrac{1}{2}N^\alpha N^\beta\Gamma^a_{bc}\big(\phi(q)\big)\nu_\alpha^b(q)\nu_\beta^c(q)+\O(|N|^3).
\end{eqnarray*}
Therefore
\begin{eqnarray*}
\partial_i\Phi^a(q,N) &=& \partial_i\phi^a(q)+N^\alpha\partial_i\nu_\alpha^a(q)+\O(|N|^2)\\
\partial_\alpha\Phi^a(q,N) &=& \big(\nu_\alpha^a-\tfrac{1}{2}N^\beta\Gamma^a_{bc}\big(\phi(\cdot)\big)\nu_\alpha^b\nu_\beta^c\big)(q)+\O(|N|^2)\\
&=& \nu_\alpha^a(q)+\O(|N|^2),
\end{eqnarray*}
where we used that $\Gamma^a_{bc}\big(\phi(q)\big)=0$ in Riemannian normal coordinates. The latter also implies that $(\overline{\nabla}_i\nu_\alpha)^a\big(\phi(q)\big)=\partial_i\nu_\alpha^a(q)$. Then the Weingarten equation~(\ref{Weingarten}) yields that
\begin{equation*}
\partial_i\nu_\alpha(q)=\omega_{i\alpha}^\gamma(q)\nu_\gamma(q)-\W_{\alpha i}^l(q)\partial_l\phi(q).
\end{equation*}
Using that $\partial_i\phi$ and $\W_{\alpha i}^l\partial_l\phi$  are tangent vectors and thus orthogonal to $\nu_\beta$ for any $i$ and $\beta$ we obtain that
\begin{eqnarray*}
\overline{g}_{ij}(q,N) &=& \big(\overline{g}_{ab}\partial_i\Phi^a\partial_j\Phi^b\big)(q,N) \\
&=&  \big(\overline{g}_{ab}\partial_i\phi^a\partial_j\phi^b-\overline{g}_{ab}\partial_i\phi^aN^\alpha\W_{\alpha j}^l\partial_l\phi^b\\
&& \quad\quad-\overline{g}_{ab}N^\alpha\W_{\alpha i}^l\partial_l\phi^a\partial_j\phi^b\big)(q)+\O(|N|^2)\\
\overline{g}_{i\beta}(q,N) &=& \big(\overline{g}_{ab}\partial_i\Phi^a\partial_\beta\Phi^b\big)(q,N)\\
&=& \big(\overline{g}_{ab}N^\alpha\omega_{i\alpha}^\gamma\nu_\gamma^a\nu_\beta^b\big)(q)+\O(|N|^2) =\overline{g}_{\beta i}(q,N)\\
\overline{g}_{\alpha\beta}(q,N) &=& \big(\overline{g}_{ab}\partial_\alpha\Phi^a\partial_\beta\Phi^b\big)(q,N)\\
&=& \big(\overline{g}_{ab}\nu_\alpha^a\nu_\beta^b\big)(q)+\O(|N|^2),
\end{eqnarray*}
Since $\phi$ is an isometric embedding, it holds $\overline{g}_{ab}\partial_i\phi^a\partial_j\phi^b=g_{ij}$. The orthonormality of the normal frame yields $\overline{g}_{ab}\nu_\alpha^a\nu_\beta^b=\delta_{\alpha\beta}$. 
Thus
\begin{eqnarray*}
\overline{g}_{ij}(q,N) 
&\stackrel{(\ref{symmetry})}{=}& g_{ij}(q)-2N^\alpha{\rm II}_{\alpha ij}(q)+\O(|N|^2),\\
\overline{g}_{i\beta}(q,n) &=& N^\alpha\omega_{i\alpha\beta}(q)+\O(|N|^2) \ \,=\ \,\overline{g}_{\beta i}(q,N),\\
\overline{g}_{\alpha\beta}(q,N) &=& \delta_{\alpha\beta}+\O(|N|^2).
\end{eqnarray*}

Inverting this matrix we end up with this proposition:

\medskip

\begin{proposition}\label{dualG}
The inverse metric tensor $\overline{g}^{ab}$ has the following form for all $q\in\C$:
\begin{equation*}
\overline{g}(q,\veps n)\;=\;\begin{pmatrix}
			1 & 0\\
			C^T & 1
                 \end{pmatrix}
		 \begin{pmatrix}
			A & 0\\
			0 & B
                 \end{pmatrix}
		 \begin{pmatrix}
			1 & C\\
			0 & 1
                 \end{pmatrix}(q,\veps n),
\end{equation*} 
where for $i,j,l,m=1,...,d$ and $\alpha,\beta,\gamma,\delta=d+1,..,d+k$
\begin{eqnarray*}
A^{ij}(q,\veps n) &=& g^{ij}(q)+ \veps\,2n^\alpha{\rm II}^{ij}_\alpha(q)+\O(\veps^2|n|^2),\\
B^{\gamma\delta}(q,\veps n) &=& \delta_{\gamma\delta}+\O(\veps^2|n|^2),\\
C_{\,i}^{\gamma}(q,\veps n) &=& -\veps\,n^\alpha\,\omega^\gamma_{i\alpha}(q)+\O(\veps^2|n|^2).
\end{eqnarray*}
Here ${\rm II}$ is the second fundamental form and $\omega$ are the coefficients of the connection on the normal bundle (see Appendix 1 for the definitions). 
\end{proposition}

We note that the error is of order $\veps^2|n|^2$ not only for small $|n|$ but globally, when the metric $\overline{g}$ is chosen properly outside a tubular neighborhood of $\C$ (see Section \ref{model} and \cite{WT}).
Then the use of the expansion is justified by the fast decay of functions from the subspaces $P_0$ and $P_\veps$ in the fibers. 


\subsection*{A3\ Transformation of measures}

Let  $\sigma_1$ be the density of the measure $d\overline{\mu}$ on $N\C$ and $\sigma_2$ be the density of the product of the measure $d\mu$ on $\C$ and the Lebesgue measure $d N$ on the fibers $N_q\C\cong\RRR^k$. Define $\rho:= \frac{\sigma_1}{\sigma_2}$ and
\[M_{\rho}:L^2(N\C,d\overline{\mu})\to L^2(N\C,d\mu\otimes dN),\,\Psi\mapsto\rho^{-\frac{1}{2}}\Psi.\]

$M_\rho$ is an isometry because for all $\Psi,\tilde\Psi\in L^2(N\C,d\mu\otimes dN)$
\[\int_{N\C} \overline{M_\rho\Psi}\,M_\rho\tilde\Psi\,d\overline{\mu}=\int_{N\C} \overline{\Psi}\,\tilde\Psi\,\rho^{-1}\,d\overline{\mu}=\int_{N\C} \overline{\Psi}\,\tilde\Psi\,d\mu\otimes dN.\]
Therefore it is clear that 
\[
M_\rho^*\Psi\;=\;\rho^{\frac{1}{2}}\Psi\,.
\]
One immediately concludes
\[M_\rho M_\rho^*\;=\;1\;=\;M_\rho^* M_\rho\]
and thus $M_\rho$ is unitary.
Now we note that $[\partial_b,\rho^{-\frac{1}{2}}]=-{\textstyle \frac{1}{2}}\,\rho^{-\frac{1}{2}}\,\partial_b\ln\rho\,$. So for $\Delta_{d\sigma_i}:=-\sigma_i^{-1}\partial_a \sigma_ig^{ab}\partial_b$ we have
\begin{eqnarray*}
\lefteqn{M_\rho^*(-\Delta_{d\sigma_1})M_\rho}\\
&& = -\rho^{\frac{1}{2}}\sigma_1^{-1}\partial_a \sigma_1\,g^{ab}\partial_b\rho^{-{\frac{1}{2}}}\\
&& = -\rho^{\frac{1}{2}}\sigma_1^{-1}\partial_a \sigma_1\rho^{-\frac{1}{2}}\big(g^{ab}\partial_b-{\textstyle \frac{1}{2}}g^{ab}(\partial_b\ln\rho)\big)
\end{eqnarray*}
On the one hand,
\begin{eqnarray*}
\lefteqn{\rho^{\frac{1}{2}}\sigma_1^{-1}\partial_a \sigma_1\,\rho^{-\frac{1}{2}}\,g^{ab}\partial_b}\\
&& =\rho\sigma_1^{-1}\partial_a \sigma_1\rho^{-1}g^{ab}\partial_b +{\textstyle \frac{1}{2}}g^{ab}(\partial_a\ln\rho)\partial_b\\
&& =\sigma_2^{-1}\partial_a \sigma_2g^{ab}\partial_b +{\textstyle \frac{1}{2}}g^{ab}(\partial_b\ln\rho)\partial_a
\end{eqnarray*}
and on the other hand,
\begin{eqnarray*}
\lefteqn{\rho^{\frac{1}{2}}\sigma_1^{-1}\partial_a \sigma_1\rho^{-\frac{1}{2}}{\textstyle \frac{1}{2}}g^{ab}\partial_b\ln\rho}\\
&& = -{\textstyle \frac{1}{4}}g^{ab}(\partial_a\ln\rho)(\partial_b\ln\rho)\\
&& \qquad +{\textstyle \frac{1}{2}}\big(\sigma_1^{-1}\partial_a \sigma_1g^{ab}\partial_b\ln\rho\big)+{\textstyle \frac{1}{2}}g^{ab}(\partial_b\ln\rho)\partial_a.
\end{eqnarray*}
Together we obtain
\begin{eqnarray*}
\lefteqn{M_\rho^*(-\Delta_{d\sigma_1})M_\rho}\\
&& = -\sigma_2^{-1}\partial_a \sigma_2g^{ab}\partial_b\\ 
&& \qquad -{\textstyle \frac{1}{4}}g^{ab}(\partial_a\ln\rho)(\partial_b\ln\rho)+{\textstyle \frac{1}{2}}\big(\sigma_1^{-1}\partial_a \sigma_1g^{ab}\partial_b\ln\rho\big)\\
&& =-\Delta_{d\sigma_2}\psi-{\textstyle \frac{1}{4}}g^{ab}(\partial_a\ln\rho)(\partial_b\ln\rho)+{\textstyle \frac{1}{2}}(\Delta_{d\sigma_1}\ln\rho).
\end{eqnarray*}

Because of $\Delta_{d\sigma_1}=\Delta_{\overline{g}}$ we have shown that
\begin{eqnarray}\label{extrapotential}
M_{\rho}^*(-\Delta_{\overline{g}})M_{\rho}
&=& -\Delta_{d\sigma_2}+V_\rho
\end{eqnarray}
with $V_\rho:=-{\textstyle \frac{1}{4}}g^{ab}(\partial_a\ln\rho)(\partial_b\ln\rho)+{\textstyle \frac{1}{2}}(\Delta_{d\sigma_1}\ln\rho)$. This formula was established many times before and we have provided its derivation for the sake of completeness, as it is the origin of the geometric potential.


\subsection*{A4\ Expansion of the Hamiltonian}

In order to deduce the formula for the effective Hamiltonian we need that $H_\veps=-\veps^2(\Delta_{\overline{g}})^\veps+V_{\rm c}(q,n)+W(q,\veps n)$ can be expanded with respect to the normal directions when operating on functions that decay fast enough. For this purpose we split up the integral over $N\C$ into an integral over the fibers $N_q\C$, isomorphic to $\RRR^k$, followed by an integration over $\C$. 

The following expansion is also the justification for the splitting of $H_\veps$ in (\ref{splitting}).

\begin{proposition}\label{expH}
If an operator $A$  satisfies 
\[\|A \langle n\rangle^l\|_{\L(\H)}\,\leq\,C_l,\quad\|\langle n\rangle^l A \|_{\L(\D(H_\veps))}\,\leq\,C_l\] 
for every $l\in\NNN$, then the operators $H_\veps A$ and $A H_\veps$ can be expanded in powers of $\veps$ on $\L\big(\D(H_\veps),\H\big)$:
\begin{eqnarray*}
H_\veps\,A &=& \big(H_0\,+\,\veps H_1\big)\,A\,+\,\O(\veps^2),\\
A\,H_\veps &=& A\,\big(H_0\,+\,\veps H_1\big)\,+\,\O(\veps^2),
\end{eqnarray*}
where $H_0$ and $H_1$ are the operators associated with 
\begin{eqnarray}\label{expcompl}
\langle\Psi|H_0\Psi\rangle &=& \int_\C\int_{\RRR^k}g^{ij}(\overline{\veps\nabla^{\rm h}_i\Psi})\,\veps\nabla^{\rm h}_j\Psi\,dn\,d\mu\,+\,\langle\Psi|H_{\rm f}\Psi\rangle,\nonumber\\
\langle\Psi|H_1\Psi\rangle &=& \int_\C\int_{\RRR^k} 2n^\alpha{\rm II}_\alpha^{ij}(\overline{\veps\nabla^{\rm h}_i\Psi})\,\veps\nabla^{\rm h}_j\Psi\\
&& \quad\qquad\qquad+n^\alpha(\partial_\alpha W)_{n=0}|\Psi|^2\,dn\,d\mu,\nonumber
\end{eqnarray}
where $\nabla^{\rm h}$ is the horizontal connection (see Section \ref{horizontal} for the definition).
\end{proposition}

To derive this let $P$ with $\|\langle n\rangle^lP\|_{\L(\D(H_\veps))}\leq C_l$ for $l\in\NNN_0$ be given. The similar case of a $P$ with  $\|P\langle n\rangle^l\|_{\L(\H)}\leq C_l$ for all $l\in\NNN_0$ will be omitted. 

\smallskip

We set $\Psi_P:=P\Psi$. By definition of $H_\veps$ it holds
\begin{eqnarray}\label{tick}
\langle\Psi\,|H_\veps\Psi_P\rangle 
&=& \big\langle\Psi\big|-\veps^2(\Delta_{\overline{g}})^\veps\Psi_P\big\rangle \nonumber\\
&& \,+\,\big\langle\Psi\big|(V_{\rm c}(q,n)+W(q,\veps n))\Psi_P\big\rangle.
\end{eqnarray}

The formula (\ref{extrapotential}) implies that
\begin{eqnarray}\label{trick}
\lefteqn{
\big\langle\Psi\,\big|-\veps^2\Delta_{\overline{g}} \Psi_P\big\rangle}\nonumber\\
&& \quad=\, \int_\C\int_{\RRR^k} \veps^2\,\overline{g}^{ab}\overline{\partial_a\Psi}\partial_b\Psi_P+\veps^2V_\rho\overline{\Psi}\Psi_P\,dN\,d\mu\nonumber\\
&& \quad=\, \int_\C\int_{\RRR^k} \veps^2\,\overline{g}^{ab}\overline{\partial_a\Psi}\partial_b\Psi_P\,dN\,d\mu
+\O(\veps^2).\nonumber
\end{eqnarray}
We emphasize once again that the remaining term may be of order $1$ for a $\Psi$ with energy of order $1$.
%
%
To calculate $-\veps^2(\Delta_{\overline{g}})^\veps$ we have to replace $N$ by $\veps n$ in the above formula. Then we may exploit $\|\langle n\rangle^2P\|_{\L(\D(H_\veps))}\leq C_2$ to insert the expansion for $\overline{g}$ from Proposition \ref{dualG} into (\ref{trick}). Noting that the rescaling $n=N/\veps$ does not change $\partial_{i}$ and replaces $\partial_{\alpha}$ by $\veps^{-1}\partial_{\alpha}$ we obtain that
\begin{eqnarray}\label{track}
\lefteqn{\big\langle\Psi\,\big|-\veps^2(\Delta_{\overline{g}})^\veps \Psi_P\big\rangle}\nonumber\\
&=& \int_{\C}\int_{\RRR^k} \overline{\big(\veps\partial_{i}+C_{\,i}^{\alpha}(q,\veps n)\partial_{\alpha}\big)\Psi}\, A^{ij}(q,\veps n) \nonumber\\
&& \qquad\qquad\qquad\qquad\qquad\quad \times\big(\veps\partial_{j}+C_{\,j}^{\beta}(q,\veps n)\partial_{\beta}\big)\Psi_P \nonumber\\
&& \qquad\quad \,+\, \overline{\big(\partial_{\alpha}\Psi\big)}\, B^{\alpha\beta}(q,\veps n)\,\partial_{\beta}\Psi_P\, dn\,d\mu\,+\,\O(\veps^2)\nonumber\\
&=& \int_{\C}\int_{\RRR^k}\big(\overline{\veps\big(\partial_{i}-n^\gamma\omega_{i\gamma}^{\alpha}(q)\partial_{\alpha}\big)\Psi}\big)\, \big(g^{ij}+\veps\,2n^\alpha{\rm II}^{ij}_\alpha\big)\nonumber\\
&& \qquad\qquad\qquad\qquad\qquad \times\,\veps\big(\partial_{i}-n^\gamma\omega_{i\gamma}^{\beta}(q)\partial_{\beta}\big)\Psi_P \nonumber\\
&& \quad\qquad\qquad\qquad \,+\, \overline{\big(\partial_{\alpha}\Psi\big)}\, \delta^{\alpha\beta}\,\partial_{\beta}\Psi_P\, dn\,d\mu\,+\,\O(\veps^2)\nonumber\\
&=& \int_{\C}\int_{\RRR^k}\big(\overline{\veps\nabla^{\rm h}_i\Psi}\big)\, \big(g^{ij}+\veps\,2n^\alpha{\rm II}^{ij}_\alpha\big) \veps\nabla^{\rm h}_{j}\Psi_P \nonumber\\
&& \quad\qquad\qquad\qquad \,+\, \overline{\Psi}\, (-\Delta_n)\Psi_P\, dn\,d\mu\,+\,\O(\veps^2),
\end{eqnarray}
where we used (\ref{horder}) and $\Delta_n=\delta^{\alpha\beta}\partial_{\alpha}\partial_{\beta}$.
Due to $\|\langle n\rangle^2P\|\leq C_2$ a Taylor expansion of $W(q,\veps n)$ in the fiber yields 
\[W(q,\veps n)P=\big(W(q,0)+\veps n^\alpha(\partial_{\alpha}W)(q,0)\big)P+\O(\veps^2).\]
Plugging this and (\ref{track}) into (\ref{tick}) we obtain the claim when we recall the definition $H_{\rm f}:=-\Delta_n+V_{\rm c}(q,n)+W(q,0)$.


\subsection*{A5\ Derivation of the effective Hamiltonian}
\ignore{
Let $\chi:\RRR\to[-1,1]$ be a Borel function with ${\rm supp}\,\chi\subset(-\infty,E]$. For any $\psi$ we set $\tilde\psi:=M_{\tilde\rho}\psi$, $\psi^\chi:= U^\veps\chi(H^\veps) U^{\veps*}\psi$, and $\tilde\psi_\chi:=U_\veps\chi(H_\veps)U_\veps^*\tilde\psi$. Of course, we have $\widetilde{\psi^\chi}=\tilde\psi_\chi$, $\|\tilde\psi\|_{L^2(\C,d\mu)}=\|\psi\|_{\H_{\rm eff}}$, and $\|\tilde\psi_\chi\|_{L^2(\C,d\mu)}\leq\|\tilde\psi\|_{L^2(\C,d\mu)}$ for all $\psi\in\H_{\rm eff}$.
 
\smallskip
 
In the following, we omit the $\veps$-scripts of $H_{{\rm eff}}^\veps,U_1^\veps$, $U_2^\veps$, and $\tilde U_\veps$ and set $\H_{\rm b}:=L^2(\C,d\mu)$. 
}
Here we derive the formula for $H_{{\rm eff}}^{(1)}=U_0H_\veps U^*_0$ stated in Theorem \ref{adiabatictheorem}. Plugging in the expansion of $H_\veps$ from the preceding appendix we have that  
\begin{eqnarray}\label{naiveexp}
H_{{\rm eff}}^{(1)} 
&=&U_0(H_0+\veps H_1)U^*_0\psi+\O(\veps^2)\nonumber\\
&=& U_0H_0U^*_0\psi+ \veps U_0H_1U^*_0\psi+\O(\veps^2)\nonumber
\end{eqnarray}

In the following we write $\langle\,\cdot\,|\,\cdot\,\rangle_{N\C}$ for the scalar product on $L^2(N\C,dNd\mu)$.  By Definition of~$U_0$ in Theorem \ref{adiabatictheorem} we have
\begin{equation}\label{null}
\langle\psi^\veps\,|\,U_0\,A\,U^*_0\psi^\veps\rangle_\C
\;=\; \langle\varphi_{\rm f}^I\psi^\veps_I|\,A\,\varphi_{\rm f}^J\psi^\veps_J\rangle_{N\C}.
\end{equation}
for any operator $A$. In view of the definition of $\nabla^{\rm h}$ in Section \ref{horizontal}, $\nabla^{\rm h}$ satisfies the usual product formula for connections:
\begin{eqnarray}\label{leibniz}
\veps\nabla^{\rm h}_i\varphi_{\rm f}^I\psi^\veps_I &=& \varphi_{\rm f}^I\veps\partial_i\psi^\veps_I+\veps\psi^\veps_I\nabla^{\rm h}_i\varphi_{\rm f}^I.
\end{eqnarray} 
We note that $\veps\nabla^{\rm h}_i\varphi_{\rm f}$ is really of order~$\veps$, while $\veps\partial_i\psi^\veps_I$ is, in general, of order $1$ due to the possibly fast oscillations of~$\psi^\veps$. Furthermore, the exponential decay of the~$\varphi_{\rm f}^I$, which implies the exponential decay of their derivatives (see \cite{WT}), guarantees that, in the following, all the fiber integrals are bounded in spite of the terms growing polynomially in $n$. 
The product formula (\ref{leibniz}) implies that
\begin{eqnarray}\label{eins} 
\lefteqn{\langle\varphi_{\rm f}^I\psi^\veps_I\,|\,H_0\,\varphi_{\rm f}^J\psi^\veps_J\rangle_{N\C}} \nonumber\\ 
&\stackrel{(\ref{expcompl})}{=}& \int_\C\int_{\RRR^k}g^{ij}\big(\overline{\veps\nabla^{\rm h}_i\varphi_{\rm f}^I\psi^\veps_I}\big)\,\veps\nabla^{\rm h}_j\varphi_{\rm f}^J\psi^\veps_J\,dn\,d\mu \nonumber
\\
&&  \quad\qquad\qquad\qquad\qquad\qquad+\,\langle \varphi_{\rm f}^I\psi^\veps_I|H_{\rm f}\varphi_{\rm f}^J\psi^\veps_J\rangle_{N\C}\nonumber\\
&=& \int_\C\int_{\RRR^k}g^{ij}\Big(\big(\overline{\varphi_{\rm f}^I\veps\partial_i\psi^\veps_I}\big)\varphi_{\rm f}^J\veps\partial_j\psi^\veps_J+\veps\big(\overline{\varphi_{\rm f}^I\veps\partial_i\psi^\veps_I}\big)\psi^\veps_J\nabla^{\rm h}_j\varphi_{\rm f}^J\nonumber\\
&&\qquad\qquad\qquad\qquad+\veps\big(\overline{\psi^\veps_I\,\nabla^{\rm h}_i\varphi_{\rm f}^I}\big)\,\varphi_{\rm f}^J\veps\partial_j\psi^\veps_J\Big)\,dn\,d\mu\nonumber\\
&& +\int_\C \langle \varphi_{\rm f}^I|H_{\rm f}\varphi_{\rm f}^J\rangle\,\overline{\psi^\veps_I}\psi^\veps_J\,d\mu +\O(\veps^2)\nonumber\\
&=& \int_\C g^{ij}\delta_{IJ}\overline{p_{{\rm eff}\,i}^{IK}\psi^\veps_K}p_{{\rm eff}\,j}^{JL}\psi^\veps_L+E_{\rm f}\delta^{IJ}\overline{\psi^\veps_I}\psi^\veps_Jdn\,d\mu+\O(\veps^2)\nonumber
\end{eqnarray}
with 
\begin{eqnarray*}
p_{{\rm eff}\,j}^{JK} &=& -\I\veps\delta^{JK}\partial_j -\veps\langle\varphi_{\rm f}^J|\I\nabla^{\rm h}_j\varphi_{\rm f}^K\rangle.
\end{eqnarray*}
Furthermore,
\begin{eqnarray}\label{zwei} 
\lefteqn{\langle\varphi_{\rm f}^I\psi^\veps_I\,|\,H_1\,\varphi_{\rm f}^J\psi^\veps_J\rangle_{N\C}} \nonumber\\ 
&\stackrel{(\ref{expcompl})}{=}& \int_\C\int_{\RRR^k}2n^\alpha{\rm II}_\alpha^{ij}\big(\overline{\veps\nabla^{\rm h}_i\varphi_{\rm f}^I\psi^\veps_I}\big)\,\veps\nabla^{\rm h}_j\varphi_{\rm f}^J\psi^\veps_J\,dn\,d\mu\nonumber\\
&& +\int_\C\int_{\RRR^k}  n^\alpha(\partial_\alpha W)_{n=0}\overline{\varphi_{\rm f}^I\psi^\veps_I}\,\varphi_{\rm f}^J\psi^\veps_J\,dn\,d\mu\nonumber\\
&=& \int_\C\int_{\RRR^k}2n^\alpha{\rm II}_\alpha^{ij}\big(\overline{\varphi_{\rm f}^I\veps\partial_i\psi^\veps_I}\big)\,\varphi_{\rm f}^J\veps\partial_j\psi^\veps_J\,dn\,d\mu\nonumber\\
&& +\int_\C  (\partial_\alpha W)_{n=0}\langle\varphi_{\rm f}^I|n^\alpha\varphi_{\rm f}^J\rangle\,\overline{\psi^\veps_I}\psi^\veps_J\,d\mu +\O(\veps)\nonumber\\
&=& \int_\C 2{\rm II}_\alpha^{ij}\langle\varphi_{\rm f}^I|n^\alpha\varphi_{\rm f}^J\rangle\,\overline{p_{\rm eff\,}^{IK}\psi^\veps_K}p_{\rm eff\,j}^{JL}\psi^\veps_L\nonumber\\
&& \qquad+(\partial_\alpha W)_{n=0}\langle\varphi_{\rm f}^I|n^\alpha\varphi_{\rm f}^J\rangle\overline{\psi^\veps_I}\psi^\veps_J\,d\mu+\O(\veps)\nonumber,
\end{eqnarray}
where we used that $p_{{\rm eff}\,j}^{JK}=-\I\veps\delta^{JK}\partial_j+\O(\veps)$. So we, indeed, obtain that
\begin{eqnarray*}
\langle\psi^\veps|H_{\rm eff}^{(1)}\psi^\veps\rangle_\C &=& \int_\mathcal{C} g_{\rm eff}^{ijIJ}\,\overline{p_{\rm eff\,i}^{IK}\psi^\veps_K}\,p_{\rm eff\,j}^{JL}\psi^\veps_L + \big(E_{\rm f}\delta^{IJ}\\ 
&&\qquad+\veps (\partial_{\alpha}W)_{n=0}\langle \varphi_{\rm f}^I | n^\alpha  \varphi_{\rm f}^J\rangle\big)\,\overline{\psi^\veps_I}\psi^\veps_J\,d\mu
\end{eqnarray*}
with
\begin{eqnarray*}
g_{\rm eff}^{ijIJ} &=& g^{ij}\delta^{IJ} \,+\, \veps \,{\rm II}^{ij}_\alpha \langle \varphi_{\rm f}^I\,|\, n^\alpha \varphi_{\rm f}^J\rangle.
\end{eqnarray*}

\smallskip


\section*{Acknowledgements}

We thank Daniel Grieser, Stefan Keppeler, David Krej$\check{{\rm c}}$i$\check{\rm r}$\'ik, Christian Loeschcke, Frank Loose,   Olaf Post, Hans-Michael Stiepan, Luca Tenuta,  Olaf Wittich, and Claus Zimmermann for helpful remarks and inspiring discussions about the topic of this paper.


\end{document}